\def\bdm{\begin{displaymath}}
\def\edm{\end{displaymath}}
\def\be{\begin{equation}}
\def\ee{\end{equation}}
\def\bea{\begin{eqnarray}}
\def\eea{\end{eqnarray}}
\def\ra{\rightarrow}
\def\infap{<\mkern -19mu{\lower1.3ex\hbox{$\sim$}\;}\,}
\def\msbar{\overline{\rm MS}\,}
\def\al{\alpha}
\def\als{\alpha_s}
\def\alst{\widetilde{\alpha}_s}
\def\bt{\beta}
\def\de{\delta}
\def\det{\widetilde{\delta}}
\def\eps{\epsilon}
\def\ga{\gamma}
\def\la{\lambda}
\def\La{\Lambda}
\def\Lams{\Lambda_{\overline{\rm MS}}}
\def\Up{\Upsilon}
\def\ga{\gamma}
\def\eul{\gamma_{\rm E}}
\def\mbbar{\overline{m}_b}
\def\mcbar{\overline{m}_c}
\def\alsgt{<\!\!\alpha_s G^2 \!\!>}
\def\qqbar{<\!\!\bar{q}q\!\!>}
\def\mup{M(\Up)}
\def\metb{M(\eta_b)}
\def\decee{\Upsilon \ra e^{+}e^{-}}
\def\decphot{ \eta_b \ra 2\gamma}
\def\gamee{\Gamma_{e^{+}e^{-}}}

\def\Vt{\widetilde{V}}
\def\Vtsi{\widetilde{V}_{\rm s.i.}}
\def\dVt{\de \widetilde{V}}
\def\Vto{\widetilde{V}^{(0)}}

\def\Uto{\widetilde{U}^{(0)}}
\def\dVsio{\de V^{(1)}_{\rm s.i.}}
\def\dVft{\de V^{(2)}_4}
\def\Lak{\Lambda(\vec{k})}
\def\vk{\vec{k}}
\def\vp{\vec{p}\,}
\def\vr{\vec{r}}
\def\vsi{\vec{\sigma}}
\def\vS{\vec{S}\,}
\def\vL{\vec{L}\,}
\documentstyle[12pt,epsf,eqsecnum]{article}
\begin{document}
\voffset= -0.4in
\baselineskip18pt
\thispagestyle{empty}
\rightline{UM--TH 93--25}
\rightline{September 16, 1993}
\vskip1.2cm
\centerline{\large{
\bf{Rigorous QCD Evaluation of Spectrum and Ground State}}}
\centerline{\large{
\bf{Properties of Heavy
$\bf q\bar{q}$ Systems; with a Precision
Determination}}}
\centerline{\large{
\bf{of {\boldmath $m_{\bf b}$}, {\boldmath $M(\eta_{\bf b})$}
.\footnote{
This work is partially supported by the U.S Department of Energy
and CICYT, Spain.}
}}}
\vskip1.2cm
\centerline{S.~Titard\footnote{E--mail address:
{\tt stephan@nantes.ft.uam.es.}}
 \,and\,F.~J.~Yndur\'{a}in}
\vskip1.2cm
\centerline{{\it Randall Laboratory of Physics}}
\centerline{{\it University of Michigan}}
\centerline{{\it Ann Arbor, MI 48109--1120}}
\centerline{and}
\centerline{{\it Departamento de F\'{\i}sica Te\'orica C-XI}}
\centerline{{\it Universidad Aut\'onoma de Madrid}}
\centerline{{\it 28049 Madrid, Spain}}
\vskip0.8cm
\centerline{\large{Abstract}}
\vskip0.8cm
We present an evaluation of heavy quarkonium states $b \bar b$,
$c\bar c$ from first principles. We use tree--level QCD (including
relativistic corrections) and the full one--loop potential;
nonperturbative effects are taken into account at the leading order
through the contribution of the gluon condensate $\alsgt$.
We use the values
 $\La({\rm 2\, loops, 4\, flavours}) = 200
\phantom{a}^{+80}_{-60}$ MeV,
 $\alsgt = 0.042 \pm 0.020
\ {\rm GeV}^4$,
but we trade the value of the quark mass by the masses of $J/\psi$,
$\Upsilon$ as input. We get good agreement in what is essentially
a zero parameter evaluation for the masses of the $1S$, $2S$, $2P$
states of $b\bar b$, the $1S$ state for $c\bar c$, and the decay
$\decee$. As outstanding result we obtain the precise determination
of $m_b$ as well as an estimate of the hyperfine splitting
$\mup-\metb$:
$\mbbar(\mbbar^2) = 4397
\phantom{a}^{+7}_{-2} \,\,(\La)
\phantom{a}^{-3}_{+4} \,\, (\alsgt)
\phantom{a}^{+16}_{-32} \,\, ({\rm systematic})$ MeV,
$\mup-\metb = 36
\phantom{a}^{+13}_{-7} \,\,(\La)
\phantom{a}^{+3}_{-6} \,\, (\alsgt)
\phantom{a}^{+11}_{-5} \,\,({\rm syst.})
$ MeV,
the first error due to that in $\La$, the second to that in
$\alsgt$ (varied independently).
For the $c$ quark, we find
$\;\mcbar(\mcbar^2) = 1306
\phantom{a}^{+21}_{-34} \,\,(\La)
\phantom{a}^{-6}_{+6} \,\, (\alsgt)$ MeV (up to systematic errors).
\setcounter{page}{0}
\newpage
\section{Introduction}
For the lowest-lying bound
states of heavy quarks ($b \bar b$, $t \bar t$
and to a certain extent, $c \bar c$) the average distance between
them is substantially smaller than the typical QCD distance,
$ r_{\Lambda}= \Lambda^{-1} \approx 1/200 $ MeV. One would therefore
expect that an analysis based on perturbative QCD, supplemented
with leading nonperturbative effects, would be applicable.
In fact, relativistic \cite{bb:ruju} and some perturbative corrections
\cite{bb:gupt,bb:buch,bb:pant} to the leading Coulombic $q \bar q$
potential,
\be
\label{eq:coulpot}
 V^{(0)}(r) = - {C_{F} \alpha_{s} \over r} \ \ , \ C_{F}= 4/3 \ ,
\ee
have been known for some time and also a number of the leading
nonperturbative corrections have been calculated \cite{bb:leut}.
Surprisingly enough, however, all existing detailed analyses include
phenomenological (and to a large extent arbitrary) confining
potentials; see for example Refs.~[4,6,7]. Now, and while this is
certainly justified for excited states, the properties of the lowest
energy
states should, as noted above, be describable rigourously with QCD.

The requisite ingredients for such an analysis (which is the
subject of the present paper) are the following. First of all, the
potential for $q \bar q$ systems has to be evaluated including
one--loop radiative corrections. This we do in Section.~2, where
we extend the evaluations of Refs.~[2,3,4] completing the spin
independent part of the potential.

The leading Coulombic potential (\ref{eq:coulpot}) gives
contributions to the quarkonium spectrum of order $\alpha_{s}^2$.
One--loop radiative corrections give contributions of order
$\alpha_{s}^3$ to the spin--independent part of the spectrum; and,
when combined with relativistic terms, contributions or order
$\alpha_{s}^4$, $\alpha_{s}^5$ are obtained. Because gluons
are massless, and the masses of light quarks (u,d,s) is negligible
it so happens that the terms of order
$\alpha_{s}^4$, $\alpha_{s}^5$ in the spin--independent piece of
the $q \bar q$ spectrum also recieve contributions from two--loop
(both) and three--loop (the last) radiative corrections.
In Section~3 we evaluate the leading two--loop correction to
terms of order $\alpha_{s}^4$, and argue that a renormalization
scheme exists in which the so--obtained terms give the quarkonium
spectrum with exact terms of orders $\alpha_{s}^2$, $\alpha_{s}^3$,
$\alpha_{s}^4$, $\alpha_{s}^5$. Moreover the scheme agrees with the
standard $\msbar$ to order $\alpha_{s}$, $\alpha_{s}^2$ and leading
log in order $\alpha_{s}^3$: given the current uncertainties in the
determination of $\Lambda$, the scheme may be considered identical
to $\msbar$ for practical purposes. Section~3 is also devoted to the
interpretation of the mass that appears in the formulas, and its
relationship to the ordinary $\msbar$ running mass $\bar m$.

Besides perturbative corrections to the Coulombic potential
account has to be taken of at least the leading non--perturbative
ones. This we do in Section~4 where we review and complete
Leutwyler's evaluations \cite{bb:leut}.

All this is applied in Sections~5, 6 for $b \bar b$ states, and in
Section~7 for
$c \bar c$ states,
 to obtain predictions for the observables that can be
evaluated rigourously within the previously developed QCD framework.
 This includes lowest hyperfine splittings, widths like
$\decee$, $\decphot$, and of course the spectrum of energy levels (as
far as the last is concerned we have found it more profitable to
evaluate quark masses from the experimentally known masses of $\Up$,
$J/\psi$).

As outstanding results we find a precision determination
of $m_b$, and a prediction for $\mup-\metb$:
\bea
\mbbar(\mbbar) & = & 4397
\phantom{a}^{+7}_{-2} \,\,(\La)
\phantom{a}^{-3}_{+4} \,\, (\alsgt)
\phantom{a}^{+16}_{-32} \,\, ({\rm syst})\;,
\\
\mup-\metb & = & 36
\phantom{a}^{+13}_{-7} \,\,(\La)
\phantom{a}^{+3}_{-6} \,\, (\alsgt)
\phantom{a}^{+11}_{-5} \,\, ({\rm syst})\;,
\eea
both in MeV, and where the errors ($\La$) are obtained by
varying
\bdm
 \La({\rm 2\, loops, 4\, flavours}) = 200
\phantom{a}^{+80}_{-60} \,\, {\rm MeV}\;,
\edm
and the ($\alsgt$) ones by taking
\bdm
 \alsgt = 0.042 \pm 0.020
\ {\rm GeV}^4
\edm
for the gluon condensate (for systematic errors see text).

The results for the decay $\decee$ as well as for the 2S--1S
and 2P--1S splittings in bottomium are less precise than the
previous ones, but the QCD analysis still yields a good
description. This is no more the case for the $c \bar c$ system
where only the mass of the c quark is given with any precision.
We get
\bdm
 \mcbar(\mcbar) =  1306
\phantom{a}^{+21}_{-34} \,\,(\La)
\phantom{a}^{-6}_{+6} \,\, (\alsgt) \
{\rm MeV} \; ,
\edm
with unknown (and probably large) systematic errors.

The article is finished in Section~8. Here we present a summary
of results (and prospects for improvement) as well as a discussion
both of the inapplicability of perturbative QCD for excited states
and the connection with phenomenological analyses.

\section{The $q \bar q$ potential to one loop, including
relativistic corrections}

The diagrams that contribute to the $q \bar q$ potential are
shown in Fig.~1. (The kinematics is shown in the tree level
diagram Fig.~2). We will discuss discuss them as we give the
pertinent results. We will first work in momentum space
\footnote{ The coordinate space potentials, $V(\vec r)$, are
related to the momentum space ones, $\widetilde{V}(\vec{k})$, by
Fourier transform:
\bdm
V(\vec{r}) = (2 \pi)^{-3} \int d^3 k \,\exp (i \vec{k}
\cdot \vec{r}) \,
\widetilde{V} (\vec{k}) \;.
\edm
}.

The tree--level potential (Fig.~2) is
\be
\Vto(\vk) = \frac{\pi C_{F} \als}{m^2} \,\Uto(\vk)
+ \Vt_{\rm kin}\, ,
\ee
where $\Vt_{\rm kin}$ is the relativistic correction to the kinetic
energy (which we conveniently consider as a perturbation),
\be
\Vt_{\rm kin} = - \frac{\vp^4}{4 m^3}\;,
\ee
and
\be
\label{eq:potUo}
\Uto(\vk) = -\frac{4 m^2}{\vk^2}-\frac{4 \vp^2}{\vk^2} +
\frac{4}{3} \vS^2 + 6 \Lak + \frac{1}{3} T(\vk)\;.
\ee
Here the spin--spin $\vS^2$, spin--orbit $\La$, and tensor T
operators are
\bea
\vS = \frac{\vsi_1 + \vsi_2}{2} \,\, & , & \,\,
\vsi_1 \cdot \vsi_2  =  2 \vS^2 -3   \,\,\, ; \nonumber
\\
\Lak & = & - i\, \frac{\vS \cdot (\vk \times \vp)}{\vk^2}
\,\,\, ;
\\
T & = & \vsi_1 \cdot \vsi_2 - \frac{3}{\vk^2}\,
(\vk \cdot \vsi_1)(\vk \cdot \vsi_2) \,\, .
\nonumber
\eea
By $m$ we denote the quark mass; the reduced mass would be
\bdm
m_{\rm red} = m/2\; .
\edm
At one--loop we have, refering to the notation of Fig.~1,
\bea
\label{eq:pota}
\Vt_{\rm a} & =  \displaystyle{\frac{C_F^2 \als^2}{m^2}} &
 \left\{
 - \frac{3 \pi^2 m}{k} + 6 - \frac{8}{3} \log \frac{\la}{m}
 - \frac{4}{3} \log \frac{k}{m}
 \right. \nonumber
\\
 &  & \left.
+ \frac{4 \pi^2 m}{3 k} \, \vS^2 + (2 \log \frac{k}{m}
 - \frac{8}{3}) \vS^2 + \frac{\pi^2 m}{12 k} \, T +
 \frac{1}{3} \, T
 \right\}
\\
 & & + \frac{C_{F}^2 \als^2}{m^2} ( 2 \log \frac{\la}{m} -
2 \log \frac{k}{m}) \, \Uto \,\, . \nonumber
\eea
$\Uto$ here and in the following given by (\ref{eq:potUo}).
Further,
\bea
\label{eq:potb}
\Vt_{\rm b} & = &
\frac{(C_{F}^2 - \frac{1}{2} C_F C_A)}{m^2}\als^2
  \left\{ \frac{\pi^2 m}{k} - 2 + 6 \log \frac{k}{m}
 + (-2 \log \frac{k}{m} +\frac{2}{3})\vS^2 - \frac{1}{3}T
  \right\} \nonumber
\\ & & +
\frac{(C_{F}^2 - \frac{1}{2} C_F C_A)}{m^2}\als^2
  \left\{
 -2 \log \frac{\la}{m} +2 \log \frac{k}{m}
  \right\} \Uto \,\, ,
\\
\label{eq:potc}
\Vt_{\rm c} & = &
\frac{(C_{F}^2 - \frac{1}{2} C_F C_A)}{m^2}\als^2
  \left\{
  -2  +\frac{4}{3} \vS^2 + \frac{1}{3}T
 + 4 \La - \frac{8}{3} \log \frac{\la}{m}
  \right\} \nonumber
\\ & & +
\frac{(C_{F}^2 - \frac{1}{2} C_F C_A)}{m^2}\als^2
  \left\{
 -\frac{1}{2} \log \frac{k^2}{\mu^2}
 + \log \frac{k}{m} +2 \log \frac{\la}{m} +2
  \right\} \Uto \,\, ;
\eea
(\ref{eq:pota}), (\ref{eq:potb})
 can be obtained from the QED calculation of Gupta et al.
\cite{bb:gupt2}, up to a colour factor. (\ref{eq:potc})
required also a slight
modification to write it in the $\msbar$ renormalization scheme.
$\mu^2$ is the corresponding mass parameter; $\la$ is a fictitious
auxilliary gluon mass to regulate the infrared singularities.

Because we want to use the $V_x$ to first order in perturbation
theory, we must subtract the iteration of the tree level diagram.
This is the graph (d) of Fig.~1, and it amounts to a term
\cite{bb:gupt,bb:gupt2}
\be
\label{eq:iter}
 -\de(\Vto,\Vto) =
 \frac{C_F^2 \als^2}{m^2}
 \left\{
 \frac{ 5 \pi^2 m}{2 k} - \frac{4 \pi^2 m}{3 k}\,\vS^2
 - \frac{\pi^2 m}{12 k}\,T
 \right\}\;.
\ee
The graphs (e) of Fig.~1 give
\bea
\label{eq:pote}
\Vt_{\rm e} & = &
 \frac{C_F C_A \als^2}{m^2}
 \left\{
 -\frac{\pi^2 m}{2 k} - 4\log\frac{k}{m} + \frac{2}{3}
 + \frac{4}{3}(\log\frac{k}{m} + \frac{1}{2}) \,\vS^2
\right. \nonumber
\\
 & &
\left.
+ 4(\log\frac{k}{m} + \frac{1}{2}) \,\La
 + (\log\frac{k}{m} + \frac{1}{2}) \,\frac{T}{3}
 \right\} \nonumber
\\
 & + &  \frac{C_F C_A \als^2}{m^2}
 \left\{
 - \frac{3}{4} \log \frac{k^2}{\mu^2}
 + \frac{3}{2} \log \frac{k}{m} +1
 \right\} \, \Uto \; .
\eea
$\Vt_{\rm e}$ coincides with the result obtained in Refs.~[2,3,4]
where, however, the spin--independent terms were not fully
evaluated. (\ref{eq:pote})
corresponds to the Dirac and Pauli form factors:
\bea
 F_{1e} & = &
 \frac{3 C_A \als}{8 \pi}
 \left(
 \log\frac{\mu^2}{m^2} + \frac{4}{3} + \frac{\pi^2 k}{6 m}
 + \frac{\vk^2}{m} \left(\log\frac{k}{m} - \frac{7}{18}
 \right)\,\,\right) \,\, ,
\nonumber
\\
 F_{2e} & = &
 \frac{C_A \als}{4 \pi}
 \left(
  2 \log\frac{k}{m} + 1
 \right)\,\, .
\nonumber
\eea
The next terms are those arising from diagrams (${\rm f}_1$) and
(${\rm f}_2$) in Fig.~1. In (${\rm f}_2$) we sum over the
quarks much lighter than the $q \bar q$; we take that they are
$n_f$ of these. Thus, for $b \bar b$, the quarks contributing to
(${\rm f}_2$) would be u, d, s and c, and we have $n_f = 4$. For
$c \bar c$, only u, d, s contribute to (${\rm f}_2$) and now
$n_f = 3$. One finds
\bea
\label{eq:potf}
 \Vt_{\rm f} = \Vt_{{\rm f}_1} + \Vt_{{\rm f}_2} & = &
 \frac{C_F C_A \als^2}{m^2}
 \left(
 -\frac{5}{12} \log\frac{k^2}{\mu^2} + \frac{31}{36}
 \right) \, \Uto
\nonumber
\\
 & & +
 \frac{C_F T_F n_f \als^2}{m^2}
 \left(
 \frac{4}{12} \log\frac{k^2}{\mu^2} - \frac{5}{9}
 \right) \, \Uto \,\,.
\eea
The contribution of the loop involving a quark
like the one in the external
legs is shown in Fig.~1(q); the annihilation channels are those
in Fig.~1(Ann a, b). One has
\bea
\label{eq:potq}
 \Vt_{\rm q} & = &
 \frac{C_F T_F \als^2}{m^2} \,\frac{-4}{15} \,\, ,
\\
 \Vt_{\rm Ann} & = &  \Vt_{\rm Ann\, a} + \Vt_{\rm Ann\, b}
 \nonumber
\\
\label{eq:potann}
 & = & \frac{C_F T_F \als^2}{m^2}
 \left\{
 -4 + 2 \log 4 + (2- \log4) \vS^2
 \right\} \,\, .
\eea
We finally consider the correction to the external legs,
Fig.~1(legs), which we define as including the wave function
$Z_\psi$ and mass $Z_m$ renormalizations. It will prove
convenient to choose $Z_\psi$ in the $\msbar$ scheme, but take
$Z_m$ {\em in the mass shell scheme}. That is to say, $m$
{\em will be taken to be the pole of the quark propagator}.
We will discuss this in more detail in Sec.~3; we now give the
result for $\Vt_{\rm legs}$:
\be
\label{eq:potlegs}
\Vt_{\rm legs} =
 \frac{C_F^2 \als^2}{m^2}
 \left\{
 \frac{1}{2} \log\frac{k^2}{\mu^2} - \log\frac{k}{m} -
 \log\frac{\la}{m} - 2
\right\} \Uto \,\,.
\ee
Adding everything, we find the potential for the $q \bar q$ system,
exact to one--loop and including relativistic corrections:
\bea
 \Vt = \Vt_{\rm s.i.} & + &
 \left\{
 (1 + \det_{\rm hf}) \frac{4}{3} \vS^2
 +  (1 + \det_{\rm LS}) \,6\, \Lak
\right.
\nonumber
\\
 & &
\ \ \ \ +\left.
 (1 + \det_{\rm hf}) \frac{1}{3} \,T
 \right\}
 \frac{\pi C_F \als}{m^2} \,\, ,
\eea
where the hyperfine, spin--orbit and tensor radiative
corrections are
\bea
 \det_{\rm hf} & = &
 \left\{
  \frac{-11 C_A + 4 T_F n_f}{6} \log \frac{k}{\mu}
 + \frac{7 C_A}{4} \log \frac{k}{m} - \frac{5}{9} T_F n_f
\right.
\nonumber
\\
 & &
\left.
 + \frac{3}{2}(1-\log 2) T_F + \frac{11 C_A - 9 C_F}{18}
\right\} \frac{\als}{\pi} \,\, ,
\nonumber
\\
\label{eq:dels}
 \det_{\rm LS} & = &
 \left\{
  \frac{-11 C_A + 4 T_F n_f}{6} \log \frac{k}{\mu}
 + \frac{2 C_A}{3}\log \frac{k}{m}
 -\frac{5}{9} T_F n_f
 + \frac{31 C_A + 24 C_F}{36}
\right\} \frac{\als}{\pi} \; ,
\\
 \det_{\rm T} & = &
 \left\{
  \frac{-11 C_A + 4 T_F n_f}{6} \log \frac{k}{\mu}
 + C_A \log\frac{k}{m}
 -\frac{5}{9} T_F n_f
 + \frac{49 C_A + 36 C_F}{36}
\right\} \frac{\als}{\pi} \; .
\nonumber
\eea
All these terms were known previously
\cite{bb:gupt,bb:buch,bb:pant};
 they are repeated here for ease of reference.

\vspace{.3in}
The spin--independent piece $\Vt_{\rm s.i.}$ is
\bea
\Vt_{\rm s.i.}(k) & = &
 \pi C_F \als(\mu^2)
 \left( -\frac{4}{\vk^2} - \frac{4 \vp^2}{m^2 \vk^2}
 \right)
\nonumber
\\
 & & \ \ \ \; \times
\left\{
 1 + \left[
 - \frac{11 C_A - 4 T_F n_f}{6} \log \frac{k}{\mu}
 + \frac{31 C_A - 20 T_F n_f}{36}
\right] \frac{\als(\mu^2)}{\pi}
\right\}
\nonumber
\\
\label{eq:potsi}
 & & + \frac{C_F \als(\mu^2)}{m^2}
 \left\{
 \frac{\pi^2 m}{2 k}(C_F- 2 C_A) +
 \frac{14 C_F - 21 C_A}{3} \log \frac{k}{m}
 \right.
\\
 & & \ \ \ \
\left.
 + \frac{-16 C_F + 4 C_A}{3} \log \frac{\la}{m} + 2 C_F
 + \frac{8}{3} C_A - \frac{64}{15} T_F + 4 T_F \log 2
 \right\}
\nonumber
\\
 & & + \Vt_{\rm kin} \, \, .
\nonumber
\eea
$\Vt_{\rm kin} = - \vp^4/4 m^2$ embodies the relativistic
corrections to the kinetic energy; as is customary and
convenient we have included them as part of the potential.

The terms of order $k^{-2}$, $k^{-1}$ in (\ref{eq:potsi})
agree with those calculated before\cite{bb:gupt,bb:pant};
the terms constant and proportional to $\log k$ are new.
(\ref{eq:potsi}) also agrees with the positronium calculation of
Ref.~\cite{bb:gupt2} after replacement
\bdm
\als \ra \al \,\, ; \,\,
 m \ra m_{\rm e} \,\, ; \,\,
 C_F,T_F \ra 1 \,\, ; \,\,
 C_A,n_f \ra 0 \,\, , \,\,
\edm
which checks some of the O($k^0$), O($\log k$) terms of
(\ref{eq:potsi}).

Eq.~(\ref{eq:potsi}) features a residual dependence on
$\log \la$, the fictitious gluon mass. This infrared divergence,
due to the neglect of binding of the quarks in the calculation,
 has the same origin as that in the Lamb shift in hydrogen and is
 likewise treated. It turns out that we have to replace
\be
\label{eq:lamb}
\log \frac{\la}{m} \ra -\frac{5}{6} - \log \bar{n} \, \, ,
\ee
where if we approximate the $q \bar q$ spectrum by a Balmer
one,
\be
\label{eq:lamb2}
\log \bar{n} \approx 0.9
\ \ \ \ \mbox{\rm for S waves.}
\ee

Eq.~(\ref{eq:potsi}) can be written in a more compact form as
\bea
\Vt_{\rm s.i.} & = &
 - \frac{\vp^4}{4 m^3} + \pi C_F \als(\mu^2)
 \left(
 -\frac{4}{\vk^2} - \frac{4 \vp^2}{m^2 \vk^2}
 \right)
\nonumber
\\
\label{eq:potsi2}
 & & \ \ \ \; \times
\left\{
 1 + \left[
 - \frac{\bt_0}{2} \log\frac{k}{\mu} + a_1
 \right] \frac{\als(\mu^2)}{\pi}
\right\}
\\
& & +
 \frac{C_F \als^2(\mu^2)}{m^2}
 \left\{
 a_2 \frac{\pi^2 m}{k} + a_3 \log \frac{k}{m}
 + a_4 \log \frac{\la}{m} + a_5
 \right\} \, \, ,
\nonumber
\eea
where
\bea
\bt_0  &=& \frac{11 C_A - 4 T_F n_f}{3} = 8.33
\,\, ; \,\,
\bt_1 = 102 - \frac{38 n_f}{3} = 51.3
\,\, ; \,\,
\nonumber
\\
 a_1 &=& \frac{31 C_A - 20 T_F n_f}{36} = 1.47
\,\, ; \,\,
 a_2 = \frac{C_F - 2 C_A}{2} = - 2.33
\,\, ; \,\,
\nonumber
\\
\label{eq:nume}
 a_3 &=& \frac{14 C_F - 21 C_A}{3} = -14.78
\,\, ; \,\,
 a_4 = \frac{-16 C_F + 4 C_A}{3} = -3.11
\,\, ; \,\,
\\
 a_5 &=& 2 C_F + \frac{8 C_A}{3} - \frac{64 T_F}{15} + 4 T_F \log 2
 = 9.92
\,\, ; \,\,
\nonumber
\\
 B &=& \frac{3}{2}(1 -\log 2) T_F - \frac{5}{9} T_F n_f
 + \frac{11 C_A - 9 C_F}{18} = 0.29
\,\, .
\nonumber
\eea
All numerical values are for $n_f = 4$. We have also given in
(\ref{eq:nume}) the values of the constants $\bt_1$, $B$ that
will appear subsequently.

For the hyperfine, LS and tensor pieces,
\bea
\label{eq:pothf}
 \Vt_{\rm hf} & = &
 \frac{4 \pi C_F \als(\mu^2)}{3 m^2} \, \vS^2
 \left\{
 1 + \left[
 - \frac{\bt_0}{2} \log\frac{k}{\mu}
 + \frac{21}{4} \log \frac{k}{m} + B
 \right] \frac{\als}{\pi}
\right\}
 \,\, ;
\\
\label{eq:potLS}
 \Vt_{\rm LS} & = &
 \frac{6 \pi C_F \als(\mu^2)}{m^2}
\, \Lak
 \left\{
 1 + \left[
 - \frac{\bt_0}{2} \log\frac{k}{\mu}
 + 2 \log \frac{k}{m} + \frac{125}{36}
 - \frac{5 n_f}{18}
 \right] \frac{\als}{\pi}
\right\}
 \,\, ;
\\
\label{eq:potT}
 \Vt_{\rm T} & = &
 \frac{\pi C_F \als(\mu^2)}{m^2} \frac{1}{3} T(\vk)
 \left\{
 1 + \left[
 - \frac{\bt_0}{2} \log\frac{k}{\mu}
 + 3 \log \frac{k}{m} + \frac{65}{12}
 - \frac{5 n_f}{18}
 \right] \frac{\als}{\pi}
\right\} \,\, .
\eea

The corresponding potentials in x--space may be found by
using the Fourier transforms collected in Appendix~I.
One gets this way the potential
\bea
\label{eq:xpotsi}
 V_{\rm s.i.}(r) & = &
V^{(0)}_{\rm eff}(r)
 + \de V^{(1)}_{\rm s.i.}(r)  \, \, ,
\\
\label{eq:xpoteff}
V^{(0)}_{\rm eff}(r) & = &
 - \frac{C_F \als(\mu^2)
 \left[
 1+ (a_1+ \eul \bt_0 / 2) \als(\mu^2)/\pi
\right]}{r} \;,
\\
\label{eq:xpotcorr1}
 \de V^{(1)}_{\rm s.i.}(r) & = &
 - \frac{ C_F \bt_0 \als^2(\mu^2)}{2 \pi}
 \frac{\log r\mu}{r}
\\
& &
  +\frac{C_F \als(\mu^2)
 \left[
 1+ (a_1+ \eul \bt_0 / 2) \als(\mu^2)/\pi
\right]}{m^2 r} \, \Delta
\nonumber
\\
 & & + \frac{ C_F \bt_0 \als(\mu^2)}{\pi m^2}
 \frac{\log r\mu}{r} \,\Delta
+ \frac{ C_F a_2 \als^2(\mu^2)}{2 m r^2}
\nonumber
\\
& &
+ \frac{ C_F a_3 \als^2(\mu^2)}{m^2}
 \left[
 -\frac{1}{4 \pi} {\rm reg}\frac{1}{r^3}
 -(\log m)
\de(\vec{r})
\right]
\nonumber
\\
& & + \frac{
 \left[
 a_5 - (5/6 + \log \bar{n}) a_4
\right]
 C_F \als^2(\mu^2)}{m^2}
\,\de(\vec{r})
\nonumber
\\
 & &
 - \frac{1}{4 m^3} \,\Delta \;.
\nonumber
\eea
$V_{\rm s.i.}$ in (\ref{eq:xpotsi}) gives the full potential
including leading relativistic corrections, and one--loop
 radiative corrections mostly included in the term
 $\de V^{(1)}_{\rm s.i.}$; but a piece of the last has been
 included in $V^{(0)}_{\rm eff}$ because it represents a
modification to the Coulomb potential that can be solved
{\em exactly} (the terms in $\de V^{(1)}_{\rm s.i.}$ will
be treated to first order in perturbation theory).
For this reason it is convenient to rewrite
$V^{(0)}_{\rm eff}$ as
\bea
\label{eq:xpoteff2}
V^{(0)}_{\rm eff}(r) & = &
 - \frac{C_F \alst(\mu^2)}{r} \,\, ,
\\
\alst(\mu^2) & = &
 \left[ 1 + ( a_1 + \eul \bt_0/2)
 \frac{\als(\mu^2)}{\pi}
 \right] \als(\mu^2) \, .
\eea

\section{ Two and Three Loop Corrections. The Full
perturbative Potential. Connection with the
 $\overline{\bf MS}$ Scheme.}

\subsection{Two and Three--loop Corrections.}

Using the potential of e.g. Eq.~(\ref{eq:xpotsi}) one gets
 the mass $M$ of a quarkonium state in terms of the quark mass
$m$ by a formula of the type
\be
\label{eq:mass}
 \frac{1}{2}M = m
 \left\{
 1 + c_2 \als^2
+ c_3 \als^3
+ c_4 \als^4
+ c_5 \als^5
 + \ldots
 \right\} \, \, ;
\ee
we neglect for the moment nonperturbative effects.
The term $c_2$ is obtained by solving exactly with
$V^{(0)}_{\rm eff}$ (Balmer term, $c_2 = - \alst^2 C_F^2/8$);
 the remaining terms are found as expectation values of
 $\de V^{(1)}_{\rm s.i.}$. Because each power $1/r$ is equivalent
to $1/a= (1/2) m C_F \alst$ ($a$ is the equivalent of the Bohr radius)
it follows that the terms in $\de V^{(1)}_{\rm s.i.}$,
Eq.~(\ref{eq:xpotcorr1}),
 although nominally of order $\als$ or $\als^2$, or even $\als^0$
for the last term there, yield powers $\als^3$, $\als^4$, $\als^5$.

In QED we could have renormalized in the mass shell scheme, and all
 three coefficients $c_3$, $c_4$, $c_5$ would be given exactly by the
expectation values of the
$\dVsio$ in~(\ref{eq:xpotcorr1}).
In QCD the situation is more complicated. Because gluons are
massless, and we neglect the mass of light quarks, it so happens
that two--loop radiative corrections contribute\footnote{
Note that this is only true for the spin--independent piece. Spin
dependent effects of order $\als^4$, $\als^5$ are given exactly by
relativistic and one--loop radiative effects.}
to $c_4$ and $c_5$; and three--loop ones contribute to $c_5$: the
only terms known exactly are $c_2$ and $c_3$.

Failing a detailed calculation of the two and three--loop
corrections to the potential we can use the following trick to
be able to still use the information contained in $c_4$, $c_5$ as
obtained from $\dVsio$. One notes that the leading correction (in
terms of $\log \mu^2$) to $c_4$ may be obtained by use of the
renormalization group. To do this one replaces, in the tree level
potential (in momentum space), i.e $- 4 \pi C_F \als/\vk^2$,
the constant $\als$ by $\als(\vk^2)$. To two loops,
\be
\als(\vk^2) = \frac{2 \pi}{\bt_0 \log k/\La}
 \left\{
 1 - \frac{\bt_1}{\bt_0^2} \frac{\log
(\log \vk^2/\La^2)}
{\log \vk^2/\La^2}
\right\} \; ,
\ee
$\bt_0$, $\bt_1$ given in (\ref{eq:nume}). One then writes
\bdm
 \log k/\La = \log k/\mu  + \log \mu/\La  \, \, ,
\edm
and expand in inverse powers of $\log \mu/\La$. In this
way one finds, first, a term in $\bt_0 \log k/\mu$ which on
inspection is seen to be already included in $\dVsio$,
Eq.~(\ref{eq:xpotcorr1}), and the desired leading two--loop
correction to be added to $\Vtsi$ in~(\ref{eq:potsi2}):
\be
\label{eq:potcorr2}
 \dVt_4^{(2)}
= \frac{- 4 \pi C_F \als(\mu^2)}{\vk^2}
 \left[
 \frac{\bt_0^2}{4}  \log^2 \frac{k}{\mu}
 - \frac{\bt_1}{8 \bt_0} \log\frac{k}{\mu} +{\rm const.}
\right] \frac{\als^2(\mu^2)}{\pi^2} \,\, ,
\ee
and the contribution of $\dVt_4^{(2)}$, $c_4^{(2)} \als^4$, when
added to $c_4 \als^4$, yields exactly the dominant part
(in $\log\mu$). Because of this it follows that a
renormalization scheme exists in which
\be
(c_4 + c_4^{(2)}) \als^4 = {\rm exact\ O}(\als^4)\ {\rm term.}
\ee
This scheme agrees with the standard $\msbar$ scheme to one--loop
and hence the corresponding parameter $\La$ differs from the
$\msbar$ one, $\Lams$, only in terms of order $\als$
\be
 \La = \left\{
 1 + \rho_1 \frac{\als}{\pi}
\right\} \Lams \, .
\ee
Writing further
\be
 \La = \left\{
 1 +
\rho_1 \frac{\als}{\pi}
+ \rho_2(\mu) \frac{\als^2}{\pi^2}
\right\} \Lams \,\, ,
\ee
and adjusting $\rho_2(\mu)$ we can get that, for a particular
case (say, ground state) of the mass formula (\ref{eq:mass}),
also the term $c_5 \als^5$ obtained with $\dVsio$ agree with the
exact O$(\als^5)$ term.

It should be clear that we are somehow trading the lack of
knowledge of $c_4$, $c_5$ by lack of knowledge of $\La$.
What we gain is that we may use the size of the terms
$(c_4 + c_4^{(2)}) \als^4$, $c_5 \als^5$, as a control of the
convergence of the perturbative series: what we lose in
knowledge of $\La$ should not matter much, as the errors
in our knowledge of $\La$ ($\approx (1/2) \La$) are likely larger
than the error comitted by replacing our $\La$ by $\Lams$.

\subsection{ The full potential.}

In our scheme, then, we have the full potential
\bea
\label{eq:xpotfull}
 V(r) & = &
V^{(0)}_{\rm eff}(r)
 + \de V^{(1)}_{\rm s.i.}(r)
 + \de V^{(2)}_4(r)  \, \, ,
\eea
where $V^{(0)}_{\rm eff}(r)$,
 $\dVsio(r)$ are given in
Eqs.(\ref{eq:xpotsi}), (\ref{eq:xpoteff}), (\ref{eq:xpotcorr1})
and $\dVft$ is the Fourier transform
of~(\ref{eq:potcorr2}):
\bea
 \dVft(r) & = &
 -\frac{C_F \bt_0^2 \als^3}{4 \pi^2}
  \frac{\log^2 \mu r}{r}
 - \left[
 \frac{\eul \bt_0^2}{2 \pi^2} + \frac{\bt_1}{2 \pi \bt_0}
 \right] C_F \als^3   \frac{\log \mu r}{r}
\nonumber
\\
\label{eq:xpotcorr2}
& & \ \ \
 - \left[
 \frac{\eul^2 \bt_0^2}{4 \pi^2}
+ \frac{\bt_0^2}{48}
+ \frac{\eul \bt_1}{2 \pi \bt_0} + {\rm const}
 \right] C_F \als^3   \frac{1}{r} \; .
\eea
Here $\als= \als(\mu^2)$, and the constant is the same as
in Eq.~(\ref{eq:potcorr2}), and we recall again that only the
first term, proportional to $\log^2\mu r$, agrees exactly with
the $\msbar$ scheme one.

For future reference we give here also the LS,
hyperfine and tensor interactions in x--space. From
Eqs.~(\ref{eq:pothf}), (\ref{eq:potLS}), (\ref{eq:potT})
and Appendix~I,
\bea
\label{eq:xpotLS}
 V_{\rm LS}(\vr) & = &
 \frac{3 C_F \als}{2 m^2 r^3} \vS \cdot \vL
 \left\{
 1 + \left[
  \frac{\bt_0}{2}( \log r\mu -1)
 \right.
 \right.
\\
 & & \ \ \ \
\left.
\left.
 + 2 \left(1 - \log m r + \frac{125 - 10 n_f}{36}
 \right)
  \right] \frac{\als(\mu^2)}{\pi}
  \right\} ,
\nonumber
\\
\label{eq:xpothf}
 V_{\rm hf}(\vr) & = &
\frac{4 \pi C_F \als(\mu^2)}{3 m^2} \vS^2
\, \left\{
 \de(\vr) + \left[
 \frac{\bt_0}{2}
 \left(
 \frac{1}{4 \pi} {\rm reg}\frac{1}{r^3}
 + (\log\mu) \de(\vr)
 \right)
\right.
\right.
\\
 & & \ \ \ \
 \left.
 \left.
 -\frac{21}{4} \left(
 \frac{1}{4 \pi} {\rm reg}\frac{1}{r^3}
 + (\log m + B) \de(\vr)
 \right) \right] \frac{\als(\mu^2)}{\pi}
 \right\} \,\, ,
\nonumber
\\
\label{eq:xpotT}
 V_{\rm T}(\vr) & = &
 \frac{C_F \als}{4 m^2 r^3} S_{12}
 \left\{
 1 + \left[
  D + \frac{\bt_0}{2}\log r\mu   -3 \log m r
  \right] \frac{\als(\mu^2)}{\pi}
  \right\}\, \, ,
\\
 D & = &
 \frac{4}{3}\left(-\frac{\bt_0}{2} + 3\right) +\frac{65}{12}
 - \frac{5 n_f}{18} \;=\; 2.75 \;(n_f = 4) \,\, ,
\nonumber
\\
 S_{12} & = & \frac{3}{r^2}
 (\vr \cdot \vsi_1)
 (\vr \cdot \vsi_2)
 -  \vsi_1 \cdot \vsi_2  \;=\; 2\left(
\frac{3}{r^2}r_i r_j -\de_{ij}
 \right)
 S_i S_j  \,\, ;
\nonumber
\eea
$B$ given in Eq.~(\ref{eq:nume}).

A last point to be discussed in this subsection is the
expression for $\als(\mu^2)$. We will use either a two--loop
expression,
\bea
 \label{eq:alstl}
 & & \als(\mu^2) =
 \frac{2 \pi}{\bt_0 \log \mu/\La}
 \left\{
 1 - \frac{\bt_1}{\bt_0^2}
\frac{\log(\log \mu^2/\La^2)}{\log \mu^2/\La^2}
 \right\}  \,\, ,
 \\
 && \La({\rm 2\, loops}, n_f =4) =  200
\phantom{a}^{+80}_{-60} \,\, {\rm MeV}
  \,\, ,
\nonumber
\eea
or a three--loop one,
\bea
 \label{eq:alsthl}
 \als(\mu^2) & = &
 \frac{2 \pi}{\bt_0 \log \mu/\La}
 \left\{
 1 - \frac{\bt_1}{\bt_0^2}
\frac{\log(\log \mu^2/\La^2)}{\log \mu^2/\La^2}
 \right.
 \\
 & & \ \ \ \
+ \left.
 \frac{\bt_1^2
\log^2(\log \mu^2/\La^2)
 -\bt_1^2 \log(\log \mu^2/\La^2) - \bt_1^2 +\bt_2\bt_0}
 {\bt_0^4 \log^2 \mu^2/\La^2}
 \right\} \,\, ,
\nonumber
\eea
where
\bdm
\bt_2 \simeq 406.4 \;{\rm for}\; n_f=4
\edm
and now we choose
\bdm
 \La({\rm 3\, loops}, n_f =4) = 185
\phantom{a}^{+80}_{-60} \,\, {\rm MeV}
  \,\, .
\edm
Both expressions of $\als$ agree for $\mu^2 \approx m_c^2
\approx (1.4\, {\rm GeV})^2$. The values of $\La$ are taken
from e.g. Refs.~\cite{bb:yndu,bb:alta,bb:data}.

\subsection{ Connection between our mass
and the $\overline{\bf MS}$ mass.}

To complete the connection between our renormalization scheme
and the $\msbar$ one we have to relate the pole mass that appears
 e.g. in Eqs.~(\ref{eq:xpotcorr2}), (\ref{eq:xpotLS}) to the usual,
running $\msbar$ mass, $\bar{m}(\mu^2)$. Ideally,
one would take the  $\bar{m}(\mu^2)$ as obtained from e.g.
SVZ~sum rules \cite{bb:shif}, say the Gasser--Leutwyler values
\cite{bb:gass},
\bea
\label{eq:massrun}
 \mbbar(\mbbar^2) & = & 4250 \pm 100 \ {\rm MeV},
\\
 \mcbar(\mcbar^2) & = & 1270 \pm 50 \ {\rm MeV}.
\nonumber
\eea
Then from these we would
obtain $m_b$, $m_c$ to be plugged in mass formulas.
This is accomplished using the relation between both definitions of
masses (Ref.~\cite{bb:coqu} to one loop, Ref.~\cite{bb:gray} to two
loops):
\be
\label{eq:polemass}
 m_{\rm pole} \equiv m = \bar{m}(\bar{m}^2)
 \left\{
 1 + \frac{C_F \als(m^2)}{\pi} + (K - 2 C_F)
 \frac{\als^2(m^2)}{\pi^2}
 \right\} \, \, ,
\ee
where\cite{bb:gray}
\be
\label{eq:Kval}
 K \approx \left\{ \begin{array}{ll}
   14.0 & \mbox{(c--quark)} \\
   13.4 & \mbox{(b--quark)}
      \end{array}
  \right. \; .
\ee
In practice we will go in the opposite direction: we will
{\em deduce} the $m$, and hence the $\bar{m}(\bar{m}^2)$
from experimentally known quarkonium masses. This is because, first,
the errors in Eq.~(\ref{eq:massrun}) are certainly underestimated;
and secondly, we will be able to get an estimate much more reliable
than (\ref{eq:massrun}) for $m_b$.

\section{Nonperturbative Contributions}

Besides the radiative, perturbative corrections to the quarkonium
properties, a full analysis has to take into account nonperturbative
corrections. At short enough distances these may be treated to
leading order and incorporated through inclusions of the effects
due to the nonvanishing  of quark and gluon condensates,
$\qqbar$, $\alsgt$ where the expectation value is taken in the
physical (as opposed to the perturbation theoretic) vacuum.

The contribution of quark condensates to quarkonium is negligible
for heavy quarks; but that of $\alsgt$ is very important. To
evaluate it we consider the graphs of Fig.~3, where the dots at the
end of the gluon lines mean replacement of the perturbative gluon
propagator, $< T B_\mu B_\nu >$, by the nonperturbative one,
\bdm
 <:\!B_\mu B_\nu\!:> \,\,\approx \,\,
 <{\rm vac} \vert : \sum_a G_a^{\mu\nu}(0)G_{a \mu\nu}(0)
  :\vert {\rm vac} > \,\,\equiv\,\, <\! G^2 \!>.
\edm
We can split the diagrams into two sets: those that involve gluon
self--couplings (diagrams~(A), (B) in Fig.~3), and those that do not.

It so happens that the first give subleading contributions, in the
nonrelativistic limit. To show this we evaluate for example graph~(A)
in Fig.~3, which can be interpreted as the correction to the gluon
propagator due to its interaction with the physical vacuum. After
a standard calculation we find the correction
\be
\label{eq:propcorr}
 \frac{ - i g_{\mu \nu}}{k^2}
\ra
 \frac{ - i g_{\mu \nu}}{k^2}
+  \frac{ - i g_{\mu \nu}}{k^2} \frac{11 \pi \alsgt}{6 k^4}
 \,\, ,
\ee
up to terms proportional to $k_\mu$, $k_\nu$ which do not
contribute to the final result. Eq.~(\ref{eq:propcorr})
holds in the Feynman gauge. The corresponding correction
to the Coulombic potential is
\bdm
 -\frac{C_F \als}{r}
 \ra
 -\frac{C_F \als}{r}
 - \frac{11 \pi \alsgt}{144} C_F \als r^3 \,\, ,
\edm
and the ensuing shift in the energy levels becomes
\bea
\label{eq:dEnpp}
\de_{\rm NP,prop}E_{nl} & = &
 - \frac{11 \pi \alsgt}{144} C_F \als < r^3 >_{nl}
\\
 & = &
 - \frac{11 n^2 \left[
 35(n^2-1)n^2 - 30 n^2(l+2)(l-1) + 3(l+3)(l^2-1)
 \right]}{144}
\nonumber
\\
 & & \ \ \ \
 \times \,\,
m \frac{\pi \alsgt}{(m C_F \als)^4} (C_F \als)^2 \,.
\nonumber
\eea
As we will see this is suppressed as compared with
 the rest of the corrections
by the factor $(C_F \als)^2$.

We now turn to the evaluation of the graphs~C in Fig.~3
which will yield the leading nonperturbative
contributions. This calculation is simplified by noting
that, since we are interested in the nonrelativistic limit, we
may consider a standard, second--order, perturbation of
the Schr\"{o}dinger equation,
\bdm
\left(
 -\frac{1}{m}\Delta - \frac{C_F \alst}{r}
 \right)
\Psi_{nl}^{(0)}
 =
  E_n^{(0)}\Psi_{nl}^{(0)} \,\, ,
\edm
by the perturbation (chromo--electric)
\bdm
 H_{\rm c.e.} = - g \vr \cdot \sum_a t^a\vec{E}_a \,\, ,
 \, \, E_a^i = G_a^{0i} \, .
\edm
A chromomagnetic contribution involving
\bdm
 H_a^i = \eps^{ijk} G_a^{jk}
\edm
can be neglected in the nonrelativistic limit because its effects
are proportional to the {\em velocity} of the quarks. Moreover,
we may consider $\vec{E}_a$ to be constant inside quarkonium as,
for heavy quarks, its size is much smaller than $\La^{-1}$.

The calculation of the shifts of the spectrum, and wave function,
are then similar to those in second order Stark effect: we take it
that
\bdm
 <\! \vec{E}_a \!>_{\rm vac} = 0 \,\, , \,\,
 <\! E_a^i E_a^i\!>_{\rm vac} =
 - \frac{\de_{ij} \de_{ab}}{3(n_c^2-1)} \frac{<\! G^2 \!>}{4}\;.
\edm
The details of the calculation (which is due to
Leutwyler) may be found in Refs~\cite{bb:leut,bb:yndu}. One
finds the energy shifts
\bea
\label{eq:dEnp}
 & & \de_{\rm NP}E_{nl} = m \frac{\pi \eps_{nl} n^6 \alsgt}
{(m C_F \alst)^4}
 \,\, ,
\\
 & &
\eps_{10} = 1.468 \, , \,
\eps_{20} = 1.585 \, , \,
\eps_{21} = 0.998 \, , \,
\eea
(for the general expression of the $\eps$'s,
cf. Ref~\cite{bb:leut}).
We use standard spectroscopic notation.

The corresponding
correction to the wave function of the ground state
is evaluated with similar methods. The result is
 \cite{bb:leut}
\bea
 \Psi_{10}(\vr) & = &
 \left(1+ \de_{\rm NP}(r)\right)
 \Psi_{10}^{(0)}(\vr) \,\, ,
\\
\label{eq:dwfnp}
 \de_{\rm NP}(r) & = & \frac{\pi \alsgt}{3825 m^4 (C_F \alst)^6}
 \left\{
 26712 - 936 \rho^2 -156 \rho^3 -17 \rho^4
 \right\}\, ,
\\
 \rho & = & m C_F \alst r\, .
\nonumber
\eea
Finally, the hyperfine splittings also get a nonperturbative
correction: one has\cite{bb:leut}
\be
\label{eq:dEnphf}
\de_{\rm NP}^{\rm hf}E_{10} = \frac{1}{3}
 ( C_F^4 \als \alst^3) \frac{18.3 \pi \alsgt}{m^4 (C_F
\alst)^6} \, .
\ee
Note that in Eqs.~(\ref{eq:dEnp}), (\ref{eq:dwfnp}),
(\ref{eq:dEnphf}),
we have already taken into account that
the effective Coulombic potential is $V_{\rm eff}^{(0)}$,
\bdm
 V_{\rm eff}^{(0)} = - \frac{C_F \alst}{r} \, .
\edm

As one sees in Eq.~(\ref{eq:dEnp}) the nonperturbative
corrections grow with the principal quantum number as $n^6$.
This means that only for $n=1$ and, in some favourable cases,
 $n=2$, will the calculation make sense. For $n \ge 3$ the
nonperturbative effects will be much larger than the Coulombic
energies, and straight QCD will no more be applicable. This
is of course to be expected: for large $n$ the wave function
extends well into the confinement region.

The value of the gluon condensate that enters into the leading
nonperturbative effects is known from e.g. SVZ sum rules.
A review of various determinations may be found in
Refs.~\cite{bb:yndu,bb:gass}. We will take
\be
\label{eq:alsgt}
 \alsgt = 0.042 \pm 0.020 \,\,{\rm GeV}^4 \,\, ,
\ee
with generous error estimates.

\section{Energy levels and Decay rates: Formulas.}

\subsection{Energy levels}

As a first approximation we solve exactly the Schr\"{o}dinger
equation,
\be
\label{eq:schr0}
 H^{(0)} \Psi^{(0)}_{nl} = E_n^{(0)} \Psi^{(0)}_{nl} \,\, ,
\ee
with
\be
\label{eq:H0}
 H^{(0)} = - \frac{1}{m} \Delta + V^{(0)}_{\rm eff}(r)\,\, ,
\ee
and $V^{(0)}_{\rm eff}$ as given in e.g. Eq.~(\ref{eq:xpoteff2}).
This gives the Balmer spectrum,
\bea
M^{(0)}(n,l,s) & = &
 2\,m \left\{ 1 - \frac{C_F^2 \alst^2}{8 n^2}
 \right\} \,\, ,
\\
\nonumber
 \Psi^{(0)}_{nl} & = & Y_{M}^l \,R_{nl}^{(0)}\,\,.
\eea
$R_{nl}^{(0)}$ are the radial wave functions; for example, for
$n=1$, $l=0$,
\be
\label{eq:wf10}
 \Psi^{(0)}_{nl}(\vr) =
 \frac{1}{(4 \pi)^{1/2}} \frac{2}{a^{3/2}} e^{-r/a} \,\, ,
 \,\, a \equiv \frac{2}{m C_F \alst} \,\, .
\ee

Adding the rest of the terms in Eqs.~(\ref{eq:xpotfull}),
(\ref{eq:xpotcorr2}), (\ref{eq:xpotcorr1})
as perturbations, and including the leading nonperturbative
contributions (\ref{eq:dEnp})
we find the corrected spectrum that we
write as follows:
\be
\label{eq:massfull}
 M(n,l,s) \,=\, 2\, m \left\{
 1 + A_2(n) + A_3(n,l) + A_4^{(1)} + A_4^{(2)} + A_5
+ A_{\rm S} + A_{\rm NP}
 \right\}.
\ee

Here the various $A_N$ are obtained with the $\de V$, and the
formulas of Appendix II. We find,
\bea
\label{eq:a2}
 A_2(n) & = & - \frac{C_F^2 \alst^2}{8 n^2} \ ,
\\
\label{eq:a3}
 A_3(n,l)
& = &
 -\frac{C_F \bt_0 \als^2}{4 n^2 \pi m a}
 \left\{ \log \frac{n a \mu}{2} + \psi(n+l+1)
 \right\} \ ,
\\
& = &
 -\frac{C_F \bt_0 \als^2 \alst}{8 \pi n^2}
 \left\{ \log \frac{n \mu}{m C_F \als} + \psi(n+l+1)
 \right\} \ ,
\nonumber
\\
\label{eq:anp}
 A_{\rm NP}(n,l) & = &
 \frac{\pi \eps_{nl} n^6 \alsgt}{2 (m C_F \alst)^4} \ .
\eea
We will give the $A_4$, $A_5$ only for $n=1$, $l=0$. Thus,
\bea
\label{eq:a41}
 A_4^{(1)}(n=1,l=0) & = &
 - \frac{3}{16} C_F^4 \left[
 1 + \left(a_1 + \frac{\bt_0 \eul}{2} \right) \frac{\als}{\pi}
 \right] \als \alst^3
\\
 & & \ \ \ + \frac{C_F^3 a_2}{8} \als^2\alst^2
 - \frac{5 C_F^4}{128} \alst^4 \ ,
\nonumber
\\
\label{eq:a42}
 A_4^{(2)}(n=1,l=0) & = &
 - \frac{\bt_0^2 C_F^2}{16 \pi^2}
\left[
 \log \frac{\mu}{m C_F \alst} + 1 - \eul
 \right]^2 \als^3 \alst \ ,
\\
\label{eq:a5}
 A_5(n=1,l=0) & = &
 -\frac{3 C_F^4 \bt_0}{16 \pi}
 \left[ \log\frac{\mu}{m C_F \alst} - \frac{1}{3} - \eul
 \right] \als^2\alst^3
\\
 & & \ \ \
 - \frac{C_F^4 a_3}{16 \pi}
 \left[ \log\frac{1}{m C_F \alst} - 1
 \right] \als^2\alst^3
\\
 & & \ \ \
 - \frac{C_F^4
 \left[ a_5 - \left( 5/6 + \log \bar{n} \right) a_4
 \right] }{16 \pi} \als^2\alst^3 \ .
\nonumber
\eea
Note that only the leading part in $\log r$
of $\dVft$ has
been employed to calculate $A_4^{(2)}$, i.e,
\bea
\label{eq:a42l}
 A_4^{(2)} & = &
 \frac{1}{2 m} \left<
 - \frac{C_F \bt_0^2 \als^3}{4 \pi^2}
\frac{\log^2 \mu r}{r} \right>
\nonumber
\\
& & \ \ \mbox{+ terms of relative order
$\displaystyle{\frac{1}{\log \mu}}$}\;,
\nonumber
\eea
(cf. Eq.~(\ref{eq:xpotcorr2})).
Finally, $A_{\rm S}$ is obtained from the hyperfine
interaction, Eq.~(\ref{eq:xpothf}): for total spin $s$,
\bea
\label{eq:as}
A_{\rm S}(n,0,s) & = & \de_{s1}
 \frac{C_F^4 \als \alst^3}{6 n^2}
 \left\{ 1 + \left[ \frac{\bt_0}{2}
 \left( \log n - \sum_1^{n} \frac{1}{k} + \eul
 - 1 + \frac{n-1}{2 n} \right)
\right.
\right.
\nonumber
\\
& & \ \ \ \,
\left.
\left.
 - \frac{21}{4} \left(
 \log\frac{n}{C_F \alst} - \sum_{1}^{n} \frac{1}{k}
- \frac{n-1}{2 n} \right) + B \right] \frac{\als}{\pi}
 \right\}
\\
& & + A_{\rm S,NP}\,\,.
\nonumber
\eea
$A_{\rm S,NP}$ is the nonperturbative contribution to the hyperfine
splitting, Eq.~(\ref{eq:dEnphf}),
that we will neglect; it is small as compared with
(\ref{eq:anp}).
In all these Eqs.~(\ref{eq:a2}) to (\ref{eq:as})
one has to consider $\als = \als(\mu^2)$
with $\als(\mu^2)$ given by the two or three loop expressions
(\ref{eq:alstl}) and
(\ref{eq:alsthl}), and the $\alst$, $\bt_0$, $a_N$, $B$ are as in
Eqs.~(\ref{eq:nume}), (\ref{eq:xpoteff2}).

It should be realized that, when calculating the {\em
spin--independent} part of the spectrum, only the terms
$A_2$, $A_3$, $A_{\rm NP}$ and $A_{\rm S}$ are known in
the $\msbar$ scheme (the last is however subleading, and
should not be included in a strict $\msbar$ analysis). The
terms $A_4$ are only known to leading order in $\log \mu$,
and $A_5$ only in our renormalization scheme.

\subsection{Hyperfine splitting, and LS splitting.}

For $n=1$, $l=0$, (\ref{eq:as}) and (\ref{eq:dEnphf})
give for the vector--pseudoscalar splitting,
\bea
\label{eq:dmvp}
 && M(V) - M(P) = \; m \frac{C_F^4 \als \alst^3}{3}
\\
&& \ \ \ \ \times
 \left\{ 1 +
\left[
\left(\frac{\bt_0}{2} -\frac{21}{4} \right)
\left( \log\frac{\mu}{m C_F \alst} - 1 \right)
 + \frac{21}{4} \log\frac{\mu}{m} + B
\right] \frac{\als}{\pi}
+ \frac{10.2 \alsgt}{m^4 \alst^6}
 \right\}
\nonumber
\\
&& \ \ \ \ \times
 \left[
 1 + 3 \bt_0 \left(
\log\frac{\mu}{m C_F\alst} - \eul
 \right)
 \frac{\als}{4 \pi}
 \right]^2 \; .
\nonumber
\eea
The last term in square brackets is due to the perturbative
correction to the wave function at the origin. It will be discussed
in detail
in section.~5.3. Note also that $\als=\als(\mu^2)$
in (\ref{eq:dmvp}).
We can conveniently rewrite (\ref{eq:dmvp}) as
\be
\label{eq:dmvp2}
 M(V) - M(P) =
\Delta^{(0)}_{\rm hf}
 \left\{ 1 +
\de^{\rm p}_{\rm hf}
 + \de^{\rm NP}_{\rm hf}
 \right\}
 \left[ 1 +
\de_{\rm wf}
 \right]^2 \; ,
\ee
where $\Delta^{(0)}_{\rm hf}= m C_F^4 \als\alst^3/3$,
$\de^{\rm p/NP}_{\rm hf}$ refer to the
radiative/nonperturbative pieces and
$\de_{\rm wf}$ refers to the perturbative correction to the
wave function at the origin.

The radiative part of the hyperfine splittings for
$l \neq 0$, $n > 1$ can be easily evaluated. One gets
\be
\label{eq:dmhf}
M(s=1,n,l) - M(s=0,n,l)= m
\frac{C_F^4 \als^2 \alst^3}{6 \pi n^3 l(l+1)(2 l +1)}
 \left(\frac{\bt_0}{2} - \frac{21}{4} \right) +
 \mbox{Non--pert.} \, \, ,
\ee
but the nonperturbative piece, expected to be very important for
$n \geq 2$, is unfortunately not known. The same is true for the
LS and tensor splitting whose perturbative piece is given by the
expectation values
\bea
 \label{eq:dmLS}
 && < \!V_{\rm LS}\! >_{nlj} \,=\,
 \frac{3 C_F^4 \als \alst^3}{16 n^3 l(l+1)(2 l +1)}
 m \left[ j(j+1) +l(l+1) -2 \right]
 \left[ 1 + \de_{\rm wf}(n,l) \right]^2
\\
 && \ \ \
\times \; \,
 \left\{
  1 + \left[
 \left(\frac{\bt_0}{2} - 2 \right)
 \left( -1 - \psi(n+l+1) +\psi(2 l +3) +\psi(2l)
   - \frac{n -l -1/2}{n} \right)
 \right.
 \right.
\nonumber
\\
 & & \ \ \ \ \
 \left.
 \left.
 + \frac{125 - 10 n_f}{36}
 + \left(\frac{\bt_0}{2} - 2 \right) \log n
 + \frac{\bt_0}{2} \log \frac{\mu}{m C_F \alst}
 + 2 \log C_F \alst
 \right] \frac{\als}{\pi}
 \right\} \, ,
\nonumber
\\
 \label{eq:dmT}
 && < \!V_{\rm T}\! >_{nljs} \,=\,
 \frac{C_F^4 \als \alst^3}{8 n^3 l(l+1)(2 l +1)}
 m  <\! \frac{1}{2} S_{12} \!>_{ljs}
 \left[ 1 + \de_{\rm wf}(n,l) \right]^2
\\
 & & \ \ \
\times \; \,
 \left\{
  1 + \left[
 D + \left(\frac{\bt_0}{2} - 3 \right)
 \left( - \psi(n+l+1) +\psi(2 l +3) +\psi(2l)
   - \frac{n -l -1/2}{n} \right)
 \right.
 \right.
\nonumber
\\
 & & \ \ \ \ \
 \left.
 \left.
+ \frac{\bt_0}{2} \log \frac{\mu}{m C_F \alst}
+ \frac{\bt_0}{2} \left(
\log n + \eul - 1 \right)
 - 3 \log \frac{n}{C_F \alst}
 \right] \frac{\als}{\pi}
 \right\} \, ,
\nonumber
\eea
where for the states that will be of interest for us
($n= 1, 2$) there is no mixing and,
\bea
<\! \frac{1}{2} S_{12} \!>_{ljs} &=&
  \left\{
 \begin{array}{ll}
    -\frac{l+1}{2 l -1}  &,\, j=l-1 \\
     + 1                  &,\, j= l   \\
     -\frac{l}{2 l +3}  &, \, j= l+1
 \end{array}
  \right.
  \ \ \  ,\, {\rm for}\, l\neq 0  \; ;
\nonumber
\\
<\! \frac{1}{2} S_{12} \!>_{0js} &=& 0
\;\; ; \;  < \!V_{\rm LS}\! >_{n0j} \,=\, 0 \;.
\nonumber
\eea
The $\de_{\rm wf}(n,l)$ embody the perturbative
corrections to the wave function at the origin; they will
be given in Section~5.3,
eqs.~(\ref{eq:parbnl}) and (\ref{eq:dwfnl}).

\subsection {Decay rates.}

We will consider the decays $V \ra e^{+}e^{-}$,
$P \ra 2\ga$, for the $n=1,l=0$ vector and pseudoscalar states.
To leading order we have
\bea
\label{eq:gaee0}
 \Gamma^{(0)}_{e^{+}e^{-}} & = &
\frac{16 \pi (Q_q \al)^2}{M(V)^2}
 \left\vert
 \Psi^{(0)}_{10}(0)
 \right\vert^2
\, = \, 2 \left[
 \frac{Q_q^2 \al}{M(V)}
 \right]^2 (m C_F\alst)^3 \, \, ,
\\
\label{eq:ga2g0}
 \Gamma^{(0)}_{2\ga} & = &
\frac{16 \pi (Q_q^2 \al)^2}{M(P)^2}
 \left\vert
 \Psi^{(0)}_{10}(0)
 \right\vert^2
\, = \, 2 \left[
 \frac{Q_q^2 \al}{M(P)}
 \right]^2 (m C_F\alst)^3 \, \, ,
\eea
The corrections to these formulas may be split as follows. First we
have perturbative corrections to the wave function at the origin,
which we actually will consider last, as well as nonperturbative
corrections to the same quantity; these were discussed in Section~4,
and are given by Eq.~(\ref{eq:dwfnp}).
Secondly we have radiative corrections
(other than those affecting the wave function\footnote{To one loop
these corrections factorize.}).
These have been known for some time\cite{bb:barb}
and can be incorporated
multiplying the r.h.s of (\ref{eq:gaee0}), (\ref{eq:ga2g0}) by
$1 +\de_r$,
\be
\label{eq:dgar}
\de_r(e^{+}e^{-}) = - \frac{4 C_F}{\pi} \als(\mu^2)\,\, ,
\,\, \de_r(2\ga) = - \left(5 -\frac{\pi^2}{4} \right)
\frac{C_F \als(\mu^2)}{\pi}
\ee

Let us now turn to the radiative corrections to the wave
function which are also needed for the evaluations of the
hyperfine splittings, Section~5.2.
These are calculated with the Hamiltonian
\be
\label{eq:ham1}
 H^{(1)} = -\frac{1}{m} \Delta - \frac{C_F \alst}{r}
 -\frac{C_F \bt_0 \als^2}{2 \pi}\frac{\log r \mu}{r} \; ,
\ee
and we take into account only the term that is exactly
known in the $\msbar$ scheme. Note that $H^{(1)}$ would have
yielded the energy levels
\be
\label{eq:Eham1}
 2 m \left\{ 1 + A_2(n) + A_3(n,l)\right\}
\ee
(cf. Eqs.~(\ref{eq:a2}) to (\ref{eq:as})).

The term $-(C_F \bt_0 \als^2/2 \pi)r^{-1}\log r\mu\;$ could be
treated in first order perturbation theory. However, since only a
 numerical solution for $\Psi(0)$
would be available, we have found it more convenient to use a
variational method. For the ground state
 one considers the (radial) test functions
\be
\label{eq:fb}
f_b(r) = \frac{2}{b^{3/2}} e^{-r/b} \;\; ,
\ee
with $b$ a parameter to be fixed so that the expectation value
$ <\!f_b,H^{(1)} f_b \!> $ be a minimum. After an elementary
calculation one finds
\be
\label{eq:parb}
 b^{-1} = a^{-1} \left\{
 1 + \frac{\log(\mu/m C_F\alst)-\,\eul}{2 \pi} \bt_0
 \als(\mu^2) \right\} \; .
\ee

It is to be noted that, with this value of $b$,
$ <\!f_b,H^{(1)} f_b \!> $ agrees with the exact value
(\ref{eq:Eham1}) up to corrections O($\als^4$): so we
expect (\ref{eq:fb}) and (\ref{eq:parb}) to represent a
very good approximation to the wave function.

For states with higher $n,\,l$ a similar calculation
tells us that we obtain the perturbed wave functions with the
replacement
\be
\label{eq:parbnl}
 a^{-1} \ra
 b(n,l)^{-1} = a^{-1} \left\{
 1 + \frac{\log(n\mu/m C_F\alst)+\psi(n+l+1)-1}{2 \pi} \bt_0
 \als(\mu^2) \right\} \; .
\ee
We will also define
\bea
\label{eq:dwfnl}
 \de_{\rm wf}(n,l) & = &
\frac{3 \bt_0 \left(
\log (n\mu/m C_F\alst) + \psi(n+l+1)-1
 \right)}{4 \pi} \als \; ,
\\
\Psi_{nl}^{(0)} &\ra & \left( 1 + \de_{\rm wf}(n,l)\right)
\Psi_{nl}^{(0)} \;.
\nonumber
\eea

Putting now everything together
(Eqs.~(\ref{eq:dwfnp}),
(\ref{eq:dgar}),
(\ref{eq:fb}) and (\ref{eq:parb})
we find the result
\bea
 \Gamma & = & \Gamma^{(0)}
 \left( 1+ \de_r \right)
 \left\{
\left(1+\de_{\rm wf} \right)
\left(1+\de_{\rm NP} \right)
 \right\}^2 \; ,
\nonumber
\\
\label{eq:gagen}
 \de_{\rm wf} & = &
\frac{3 \bt_0 \left(
\log(\mu/m C_F\alst) - \eul
 \right)}{4 \pi} \als(\mu^2) \; ,
\\
 \de_{\rm NP} & \equiv &  \de_{\rm NP}(0) =\,
\frac{26712\, \pi \alsgt}{3825\, m^4 (C_F \alst)^6} \;,
\nonumber
\eea
and the $\Gamma^{(0)}$, $\de_r$ as given in
Eqs.~(\ref{eq:gaee0}), (\ref{eq:ga2g0}) and (\ref{eq:dgar}).

\section{Numerical Results for the $b \bar b$ System.}

\subsection{Choice of the scale $\mu$.}

Before entering into a detailed discussion a few points
should be cleared. Even for the ground state of $b \bar b$ the
average momentum is
\bdm
 \bar k \,\,=\,\, <\! \vk^2 \!>_{10}^{1/2}
\,\,=\,\, \frac{m C_F \alst}{2}
 \,\approx\, 1.24 \,{\rm GeV} \; ,
\edm
and as we will see,
\bdm
 \als \approx 0.28 \; .
\edm
This means that we are close to the region of nonapplicability
of the QCD analysis at the same time (as we will see) because
nonperturbative effects become large, and also because of the
poor convergence of the perturbative series. Roughly speaking
we have two kinds of calculation: in favourable cases, perturbative
and nonperturbative corrections are clearly smaller than the leading
terms. This happens for the mass $M(1,0)$.
In other cases, like
 the hyperfine splitting $M(V)-M(P)$,
the decays
$V \ra e^{+}e^{-}$ and $P \ra 2\ga$ and, especially, the masses
$M(2,0)$, $M(2,1)$, the corrections are almost
as large as the leading term.

In the first cases, dependence on the parameter $\mu$ (that
should be absent in an exact calculation) is slight; in situations
of the second kind, a strong dependence on $\mu$ is found. In
both cases we have the question of the choice of $\mu$: in the
first to optimize already good results; in the second to make sense
at all.

In this respect we may consider the following choices:
\begin{itemize}
  \item i) $\mu=m$. This is {\em not} a natural scale for
quarkonium, mostly dominated by $ <\! \vk^2 \!>$. Choice of
$\mu=m$ leads to very large nonperturbative effects, which of
course reflects what has just been said. We only present
results with this choice in a particularly favourable case.
  \item ii) Choose $\mu$ such that the radiative corrections
vanish.
  \item iii) Choose $\mu = \bar k$.
  \item iv) Choose $\mu$ such that the radiative corrections equal,
 in modulus, the nonperturbative ones.
This gives, generally speaking $\mu^2$ values between
$<\!\vk^2\!>$ and $m^2$.
\end{itemize}

This last choice yields the best results in the favourable cases
where on can try them all; and we in fact believe it to be
the most appropriate in all cases, as it corresponds to striking an
equilibrium between two types of effects which, for bottonium
spectroscopy are both essential.

A point has to be clarified in what respects the application
of the last criterion (iv). Best results are obtained by varying
$\mu$ when varying $\La$ within its error bars; and varying $\mu$
also when varying $\alsgt$ within its error bars. This we applied
consistently throughout all calculations.

\subsection{The ground state of $b \bar b$, and a precision
determination of $m_b$, $\mbbar$.}

The mass of the $\Upsilon$ may be calculated with the formulas
(\ref{eq:a2}) to
(\ref{eq:as}),
 taking $n=1,l=0$. As explained in the introduction, we will take
$\mup = 9.460$ GeV from experiment and {\em deduce} $m_b$, $\mbbar$.
We fisrt choose $\mu = m$. Then,
\bea
 && A_2 = -9.6 \,\;10^{-3}
\; ,\;
A_3 = - 6.1 \,\;10^{-3}
 \; , \;
\nonumber
\\
\label{eq:numA1}
 && A_4^{(1)} + A_4^{(2)} = -6.2 \,\;10^{-3}
 \; , \;
 A_5 + A_{\rm S} = 4.0 \,\;10^{-4}
 \; , \;
 A_{\rm NP} = 3.4 \,\;10^{-2}
 \; ;
\\
 && \als(m^2) = 0.19
 \; , \;
 \alst(m^2) = 0.23
 \; .
\nonumber
\eea
Moreover, and with $\La$, $\alsgt$ varying independently as
in (\ref{eq:alstl}),
 (\ref{eq:alsthl}),
 and (\ref{eq:alsgt}),
\bea
\label{eq:mb1}
 m_b & = & 4763
\phantom{a}^{+82}_{-96} \,\,(\La)
\phantom{a}^{-48}_{+46} \,\, (\alsgt) \
{\rm MeV} \; ,
\\
\label{eq:mbbar1}
\mbbar(\mbbar) & = & 4264
\phantom{a}^{+14}_{-33} \,\,(\La)
\phantom{a}^{-45}_{+42} \,\, (\alsgt) \
{\rm MeV} \; .
\eea
Convergence of (\ref{eq:numA1}) is reasonable, but the
nonperturbative piece is actually {\em larger} than even
the {\em Balmer} term.

Take now $\bar \mu$ such that criterion (iv) is fulfilled, i.e.,
such that
\bdm
 \left\vert
 A_3^{\mu = \bar{\mu}}
 \right\vert
 =
 \left\vert
 A_{\rm NP}^{\mu = \bar{\mu}}
 \right\vert \; .
\edm
We have taken $\als$ as given by a two or three loop
expression
(Eqs.~(\ref{eq:alstl})
and (\ref{eq:alsthl})): almost identical results have been obtained
in both cases. We have considered three possibilities: the one which
we consider optimum, with the whole expression
(\ref{eq:massfull}).
Then
\bea
\label{eq:mb2}
 && m_b =  4906
\phantom{a}^{+69}_{-51} \,\,(\La)
\phantom{a}^{-4}_{+4} \,\, (\alsgt) \
{\rm MeV} \; ,
\\
 \phantom{nothing}
\nonumber
\\
 && A_2 = -0.033
\; ,\;
A_3 = 2.4 \,\;10^{-3}
 \; , \;
\nonumber
\\
\label{eq:numA2}
 && A_4^{(1)} + A_4^{(2)} = - 1.0 \,\;10^{-2}
 \; , \;
 A_5 + A_{\rm S} = 2.3 \,\;10^{-3}
 \; , \;
 A_{\rm NP} = 2.4 \,\;10^{-3}
 \; ;
\\
 && \als(\bar{\mu}^2) = 0.28
 \; , \;
 \alst(\bar{\mu}^2) = 0.38
 \; , \;
 \bar{\mu} = 1.44 \phantom{a}^{+0.26}_{-0.17} \;{\rm GeV} \; .
\nonumber
\eea
Clearly,
(\ref{eq:mb2}) is compatible
with (\ref{eq:mb1}); but now the errors are much smaller, a fact
which is due to the excellent convergence of the series. In fact,
from (\ref{eq:mb2})
we see that all corrections (radiative, relativistic and
nonperturbative) are comfortably smaller than the Balmer term
$A_2$.

We have also evaluated $m_b$ keeping only
$A_2$, $A_3$, $A_4^{(1)} + A_4^{(2)}$, $A_{\rm NP}$, obtaining
\be
\label{eq:mbsyst1}
  m_b =  4917
\phantom{a}^{+82}_{-56} \,\,(\La)
\phantom{a}^{-3}_{+4} \,\, (\alsgt) \
{\rm MeV} \; ,
\ee
or just keeping the terms known exactly in the $\msbar$ scheme
$A_2$, $A_3$, $A_{\rm NP}$:
\be
\label{eq:mbsyst2}
  m_b =  4866
\phantom{a}^{+45}_{-37} \,\,(\La)
\phantom{a}^{-4}_{+4} \,\, (\alsgt) \
{\rm MeV} \; .
\ee
Other criteria give nothing essentially new. For example, if
choosing $\mu_0$ such that
$A_3 = 0$, we get
\bdm
 m_b = 4902 \ ,\ \mu_0=1.56 \ {\rm GeV} \; ,
\edm
practically indistinguishable from (\ref{eq:mb2}), but the
errors are slightly larger.
In view of this, we will accept (\ref{eq:mb2}), and use
(\ref{eq:mbsyst1}),
(\ref{eq:mbsyst2}) as giving indication of the {\em systematic}
errors. In this way we get the precise estimate
\be
\label{eq:mbfin}
  m_b =  4906
\phantom{a}^{+69}_{-51} \,\,(\La)
\phantom{a}^{-4}_{+4} \,\, (\alsgt)
\phantom{a}^{+11}_{-40} \,\, ({\rm Syst.}) \
{\rm MeV} \; .
\ee

The corresponding value of $\mbbar(\mbbar^2)$ is obtained with
Eq.~(\ref{eq:polemass}).
One finds that the dependence on $\La$ drops almost entirely
and we get the result\footnote{Compatible with
(\ref{eq:mbbar1}) but with substantially smaller errors.},
\be
\label{eq:mbbfin}
 \mbbar(\mbbar^2) =  4397
\phantom{a}^{+7}_{-2} \,\,(\La)
\phantom{a}^{-3}_{+4} \,\, (\alsgt)
\phantom{a}^{+16}_{-32} \,\, ({\rm Syst.}) \;
{\rm MeV} \; .
\ee
This last value represents a substantial refinement of the
existing estimate (\ref{eq:massrun})
\bdm
 \mbbar(\mbbar^2) =  4250 \pm 100 \ {\rm MeV}.
\edm

\subsection{Hyperfine splitting}

The mass difference $\mup -\metb$ is found
readily from
Eqs.~(\ref{eq:dmvp})
and (\ref{eq:dmvp2}).
To choose $\mu$ we  follow our standard  criterion (iv), i.e.,
select $\bar{\mu}_{\rm hf}$ such that perturbative
and nonperturbative corrections cancel each other.

\noindent
Also we fix\footnote{We do not take into account the
systematic error in $m_b$,
Eq.~(\ref{eq:mbfin}),
as it affects the hyperfine splitting by
less than 1 MeV.}
(Eq.~(\ref{eq:mbfin})),
\bdm
  m_b =  4906
\phantom{a}^{+69}_{-51}
\ {\rm MeV} \; , \;
 \La = 200
\phantom{a}^{+80}_{-60}  \ \; (\mbox{2--loop}),
\edm
and we will vary independently
\bdm
 \alsgt = 0.042 \pm 0.020 \
{\rm GeV}^4.
\edm

\noindent
With this choice $\bar{\mu}_{\rm hf}$ we find
\be
\label{eq:numupetb}
 \bar{\mu}_{\rm hf} = 2.22
\phantom{a}^{+0.53}_{-0.46} \ \, {\rm GeV},
\ \;
 \de_{\rm hf}^{\rm p}
 =  \de_{\rm wf} = - 0.22 \; , \;
 \de_{\rm hf}^{\rm NP} = 0.88 \; ,
\ee
and
\be
\label{eq:dmupetbfin}
 \mup - \metb = 36
\phantom{a}^{+13}_{-7} \,\,(\La)
\phantom{a}^{+3}_{-6} \,\, (\alsgt)
\phantom{a}^{+11}_{-5} \,\, ({\rm syst})
 \ {\rm MeV} \; ,
\ee
which is our best result.
Note that the value of $\mu$ is somewhat between the two scales
$<\!\vk^2\!>^{1/2}$, $m_b$; a phenomenon similar
to that will be encountered and
discussed in the next section~6.4.
To obtain the systematic errors we have proceded as follows.
First we fix $\mu = m C_F \alst$ and then one finds
\bdm
 \mup - \metb = 31 \;{\rm MeV}.
\edm
Alternatively, we can eliminate the wave function from the
expression for the hyperfine splitting by normalizing it to
the experimental decay $\decee$. In this way we obtain,
\bdm
 \mup - \metb = 47 \;{\rm MeV}.
\edm
Both results are compatible with
(\ref{eq:dmupetbfin}), within errors. We then take them to
furnish us with an estimate of the {\em systematic}
uncertainty, and this is what is reported in
(\ref{eq:dmupetbfin}).

One can see that perturbative and nonperturbative corrections
are substantial; but they cancel largely. Neglect of the radiative
corrections would result in an overestimate of the splitting
(as in Refs.~\cite{bb:leut,bb:gass}) to some $110$ MeV.
In Refs.~\cite{bb:gupt,bb:pant} a value of about
$35$ MeV is found, which is similar to our result; however
nonperturbative effects were
not calculated but simulated by a linear $K r$ term
in these analyses.
Thus our result,
Eq.~(\ref{eq:dmupetbfin}), is a true, precise prediction for the
mass of the $\eta_b$ particle.

\subsection{$e^{+}e^{-}$ and $2\ga$ decays.}

We will consider in detail only the $\decee$ decay; the
$\decphot$ may be best obtained as
\bea
\Gamma(\decphot) & = & Q_b^2 \,\,
 \frac{1- (5 -\pi^2/4) C_F \als/ \pi}{1 - 4 C_F \als/ \pi}
 \,\,\Gamma^{\rm exp}(\decee)
\\
 & \approx & 0.17 \;{\rm keV}\; .
\nonumber
\eea
Using Eq.~(\ref{eq:gagen}), a first approximation is found by
fixing $\bar{\mu}$ from the $\mup$ analysis of subsection~6.2.
Then, and taking the values of $m$, $\La$ from
(\ref{eq:mb2}) and (\ref{eq:numA2}),
\be
\label{eq:gamee0}
 \Gamma_{e^{+}e^{-}}^{(0)} =
 2.3 \phantom{a}^{+2.3}_{-1.0} \; {\rm keV}\
 ({\rm experimentally}\ 1.34 \pm 0.04) \; ,
\ee
a not very precise estimate.

When incorporating radiative and nonperturbative effects we are
faced with the following problem. The natural scale for
$\vert \Psi_{10}\vert$, as well as the nonperturbative corrections
is $\bar{\mu} \sim \bar{\mu}_{\rm hf} \sim \bar{k} \simeq 1.5\;
{\rm GeV}$; while for the "radiative" pieces $\de_r$ (cf.
Eq.~\ref{eq:dgar}) the natural scale is $\mu \sim m$. This last is
known from estimates of the ratios
$\Upsilon \ra \;{\rm hadrons}\;/ \decee \;$. Of course the rigorous
solution to this problem is to evaluate {\em next} order
perturbative {\em and} nonperturbative corrections, which should
take care automatically of the existence of two scales. This is not
an easy task. In the meantime we fall back on our familiar
criterion (iv): we fix $\bar{\mu}_m$ as a mean value of $\mu$
such that perturbative and nonperturbative corrections
cancel each other in $\gamee$. In this way we find,
\bea
\label{eq:gamee}
 && \gamee =
 1.01
\phantom{a}^{+0.02}_{-0.02} \,\,(\La)
\phantom{a}^{+0.18}_{-0.24} \,\, (\alsgt)
\;\;{\rm keV} \; ,
\\
\label{eq:numgamee}
&& \bar{\mu} = 2.52 \;{\rm GeV}\; ,
 \de_r^{e^{+}e^{-}} = -0.39 \; ,
 \de_{\rm wf} = -0.13 \; ,
 \de_{\rm NP} = 0.47 \; ,
\eea
(to obtain $\gamee$ the running of $\al_{\rm QED}$ was also
taken into account).
Eq.~(\ref{eq:gamee}) is certainly an improvement on
(\ref{eq:gamee0}), and it is also satisfactory that $\bar{\mu}_m$
falls halfway between the two scales,
\bdm
 \bar{k} \simeq 1.24 < \bar{\mu}_m < m\simeq 4.9 \ {\rm GeV}\;.
\edm
One peculiarity of (\ref{eq:gamee}) is that a variation of $\La$
almost produces no variation of $\gamee$. This is because $\mu$
adjusts itself so that $\mu/\La$, and hence $\als$, do not vary:
the whole error is shifted to the error due to $\alsgt$.

\subsection{Excited states ($n=2$).}

{}From Eqs.~(\ref{eq:a2}) to (\ref{eq:anp})
we get the expressions for the mass of the $2S$ and $2P$
states\footnote{Referred to in the Particle Data Tables as
$\Upsilon(2S)$ and $\chi_j(1P)$ ($j$ the total spin). For the
last we consider their {\em average} mass,
$M(2P) \equiv$ average $M(\chi_j(1P)) = 9888 \pm 27$ MeV.},
neglecting for the moment fine and hyperfine effects in
the last:
\bea
\label{eq:dm2s1s}
&&  M(2S) - M(1S) = 2 m \left\{
 \frac{3 C_F^2 \alst^2}{32}
\right.
\\
 && \ \
\left.
 + \frac{C_F^2 \bt_0 \als \alst}{32}
 \left[ 3 \log\frac{\mu}{C_F m \alst} + \frac{5}{2} -3 \eul
 - \log 2 \right] \frac{\als}{\pi}
 + \frac{(2^6 \times 1.59 - 1.47) \pi \alsgt}{2 C_F^4 m^4 \alst^4}
 \right\}
\nonumber
\\
&& \phantom{M(2S) - M(1S)} \equiv
 2m \left\{
 \de_{\rm 2S}^{\rm Balmer}
 + \de_{\rm 2S}^{\rm rad}
 + \de_{\rm 2S}^{\rm NP}
 \right\} \; ;
\nonumber
\\
\label{eq:dm2p1s}
&&  M(2P) - M(1S) = 2 m \left\{
 \frac{3 C_F^2 \alst^2}{32}
\right.
\\
 && \ \
\left.
 + \frac{C_F^2 \bt_0 \als \alst}{32}
 \left[ 3 \log\frac{\mu}{C_F m \alst} + \frac{13}{6} -3 \eul
 - \log 2 \right] \frac{\als}{\pi}
 + \frac{(2^6 \times 0.998 - 1.47) \pi \alsgt}{2 C_F^4 m^4 \alst^4}
 \right\}
\nonumber
\\
&& \phantom{M(2P) - M(1S)} \equiv
 2m \left\{
 \de_{\rm 2P}^{\rm Balmer}
 + \de_{\rm 2P}^{\rm rad}
 + \de_{\rm 2P}^{\rm NP}
 \right\} \; .
\nonumber
\eea
We have not taken into account the contributions of $A_4$, $A_5$.

If we apply our criterion, defining $\mu_{2S}$, $\mu_{2P}$ such
that $\vert \de^{\rm rad}\vert = \vert \de^{\rm NP} \vert$ then
we find, with our values for $\La$, $\alsgt$ and $m$ (note that
$\La$ has to be converted to $\La$(3 flavours)
$\simeq 250 \phantom{a}^{+80}_{-60}$ MeV),
\bea
\label{eq:dm2s1snum}
&&  M(2S) - M(1S) =
 479
\phantom{a}^{+28}_{-23} \,\,(\La)
\phantom{a}^{+49}_{-69} \,\, (\alsgt) \
\; ,
{\rm exp.}\; 563\;{\rm MeV} \; ,
\\
&& \ \
 \mu_{2S} = 0.986 \phantom{a}^{+0.287}_{-0.222} \;{\rm GeV} ,\;
 \als =0.36 \; , \alst = 0.54 \; ,
\nonumber
\\
&& \ \
 \de_{2S}^{\rm Balmer} = 0.049 \; ,
 -\de_{2S}^{\rm rad} = \de_{2S}^{\rm NP} = 0.042 \; ,
\nonumber
\eea
and
\bea
\label{eq:dm2p1snum}
&&  M(2P) - M(1S) =
 417
\phantom{a}^{+25}_{-20} \,\,(\La)
\phantom{a}^{+42}_{-59} \,\, (\alsgt) \
\; ,
{\rm exp.}\; 428\;{\rm MeV} \; ,
\\
&& \ \
 \mu_{2P} = 1.062 \phantom{a}^{+0.307}_{-0.237} \;{\rm GeV} ,\;
 \als =0.34 \; , \alst = 0.51 \; ,
\nonumber
\\
&& \ \
 \de_{2P}^{\rm Balmer} = 0.043 \; ,
 -\de_{2P}^{\rm rad} = \de_{2P}^{\rm NP} = 0.035 \; .
\nonumber
\eea

A direct calculation of the "Lamb effect" splitting,
$M(2P)-M(2S)$ is of interest. The theoretical formula is
immediate from
(\ref{eq:dm2s1snum}) and
(\ref{eq:dm2p1snum}):
\be
\label{eq:dm2p2s}
  M(2P) - M(2S) = 2 m \left\{
 \frac{C_F^2 \bt_0 \als^2\alst}{96 \pi}
 + \frac{2^6\,(1.59 - 0.998) \pi \alsgt}{2 C_F^4 m^4 \alst^4}
 \right\}\;.
\ee
We may fix here $\bar{\mu}$ by requiring
$ \vert
\de_{2P}^{\rm rad}
- \de_{2S}^{\rm rad}
 \vert = \vert
\de_{2P}^{\rm NP}
 - \de_{2S}^{\rm NP}
\vert $, or choose $\mu_{SP} = (1/2)(\mu_{2S}+ \mu_{2P})$. We get
\be
  M(2P) - M(2S) =   \left\{
    \begin{array}{ll}
      128\;{\rm MeV} &,\; \bar{\mu}= 0.805 \;{\rm GeV} \\
      214\;{\rm MeV} &,\; \mu_{SP} = 1.024 \;{\rm GeV}
    \end{array}
  \right\} \;\; ({\rm exp.:} \; 135 \;{\rm MeV})\;.
\ee
The difference between the two values is an indication of the
rather large systematic uncertainty.

As we see the agreement with experiment is
surprisingly good for the $n=2$ states, in spite of the fact that
the wave function extends into the confinement region: a fact made
apparent both by the size of $\mu \simeq 1\;{\rm GeV}$, and the
importance of the nonperturbative contributions, alsmost 90\% of the
leading ones. Obviously, it is hopeless to try to extend the
analysis to higher values of $n$: because leading nonperturbative
contributions grow like $n^6$, $n=3$ is fully outside the realm
of validity of a calculation
using perturbative and leading NP effects.

The importance of the nonperturbative effects is also seen,
{\em in absentia}, in a calculation of the LS and hyperfine
splittings in the $2P$ system. From
Eqs.~(\ref{eq:dmLS}) and
(\ref{eq:dmT}) we find (choosing $\mu = \mu_{SP}$ for the
numerical results),
\bea
\label{eq:dm2ps1s0}
&& M(2P; s=1) - M(2P; s=0)
  = \frac{C_F^4 \als^2 \alst^3 m}{288 \pi}
\left( \frac{\bt_0}{2} - \frac{21}{4} \right)
\\
&& \ \ \ \ \ \
 + (\mbox{unknown nonperturbative term} \equiv {\rm NP})
\nonumber
\\
&& \ \ \ \
\nonumber
 \simeq -0.23 \;{\rm MeV} + \,{\rm NP} \ \ \ \,
 ({\rm exp.}: \; -0.9 \pm 0.2 \;{\rm MeV})
\eea
and, for the LS and tensor splitting,
\bea
\label{eq:mj2p}
&& M_j(2P) \; = \;
 C_F^4 \als \alst^3 \,m\,
\; \left[ 1 +\de_{\rm wf}(2,1) \right]^2
\; \left[  \;\;\; \frac{j(j+1)-4}{256}
 \right.
\\
 && \ \ \
\times \; \left\{
  1+ \left[
  \left( \frac{\bt_0}{2}-2 \right)
  \left( \log \frac{\mu}{m C_F \alst} +\log 2 - \eul \right)
  + 2 \log\frac{\mu}{m} + \frac{125 -10 n_f}{36}
  \right] \frac{\als}{\pi}
  \right\}
\nonumber
\\
 && \ \ \
 \left. +\; \frac{<\!\frac{1}{2} S_{12}\!>}{384}
\; \, \left\{
  1+ \left[ D +
  \left( \frac{\bt_0}{2}-3 \right)
  \left(  1+ \log 2 - \eul +
\log \frac{\mu}{m C_F \alst} \right)
  \right] \frac{\als}{\pi}
  \right\}
 \; \; \; \; \right]
\nonumber
\\
 && \phantom{M_j(2P) \; = \;}
 \,+ (\mbox{unknown NP})\;,
\nonumber
\eea
($\de_{rm}(2,1)$ given in (\ref{eq:dwfnl})).
We get for the corresponding splittings,
\bea
\label{eq:dmj2j12p}
&& M_2(2P) - M_1(2P) =
 \left\{
 \begin{array}{ll}
  10.0\;{\rm MeV}\; (\mbox{tree level})\; ;
\\  9.1\;{\rm MeV}\; (\mbox{with radiative corrections})\; ;
\\ ({\rm exp}: \; 21 \pm 1 \;{\rm MeV})\; ;
  \end{array}
 \right.
\\
\label{eq:dmj1j02p}
&& M_1(2P) - M_0(2P) =
 \left\{
 \begin{array}{ll}
  12.5\;{\rm MeV}\; (\mbox{tree level})\; ;
\\  26.0\;{\rm MeV}\; (\mbox{with radiative corrections})\; ;
\\ ({\rm exp}: \; 32 \pm 2 \;{\rm MeV})\; ;
  \end{array}
 \right.
\eea
which, as mentioned, imply nonperturbative corrections of the
order (or slightly larger) than radiative ones.

\subsection{The $\Upsilon$ wave function.}

In some applications one needs the full wave function of a bound
state (and not only its value at $\vr=0$). Putting together the
results of Section~4, Eq.~(\ref{eq:dwfnp}),
and of Section~5.3, Eqs.~(\ref{eq:fb}) and (\ref{eq:parb}), we
can give an expression for the $\Upsilon$ wave function,
embodying leading order nonperturbative corrections and
one--loop radiative ones. We get,
\bea
\label{eq:wfup}
\Psi_{10}(\vr) &=&
 \frac{1}{(4 \pi)^{1/2}} \, f_{10}(r) \; ,
\\
f_{10}(r) &=& \frac{2}{b^{3/2}}\, e^{-r/b}
\left\{
1 + \frac{\pi \alsgt}{m^4 (C_F \alst)^6}
 \,\frac{26712 - 936 \rho^2 -156 \rho^3 -17 \rho^4}{3825}
 \right\}\; ,
\\
 b^{-1} &=& \frac{m C_F \alst}{2}
\,\left\{
 1 + \bt_0 \left[
\log\frac{\mu}{m C_F\alst}-\eul
 \right]
\,\frac{\als}{2 \pi}
\right\} \; , \;
 \rho =  2 \,\frac{r}{b} \; ,
\nonumber
\\
\alst  &=&
 \left[ 1 + \left(  \eul \bt_0/2
 + \frac{31 \,C_A - 20 \,T_F n_f}{36}
\right)
 \frac{\als}{\pi}
 \right] \, \als \; , \; \,\als \,=\, \als(\mu^2) \; .
\nonumber
\eea
This wave function is normalized to unity, with an error
\bdm
\frac{2956689216}{14630625}
\left[
\frac{\pi \alsgt}{m^4 (C_F \als)^6}
\right]^2
 \simeq  \, 0.4 \%  \;\;
({\rm for} \;\mu = 1.44 \;{\rm GeV}) \;.
\edm
Eq.~(\ref{eq:wfup}) is to be used with $\mu$ as a parameter
varying in the range given in Eqs.~(\ref{eq:numA2}),
(\ref{eq:numupetb}), (\ref{eq:numgamee}).

An effective Hamiltonian such that its expectation value in the
wave function (\ref{eq:wfup}) reproduces the ground state energy,
including radiative and nonperturbative terms (but {\em not}
relativistic corrections) is also easily seen to be
\be
\label{eq:hameff}
H_{\rm eff} =
 2\, m - \frac{1}{m} \Delta - \frac{C_F \alst}{r}
 - \frac{C_F \bt_0 \als^2}{2 \pi} \frac{\log r \mu}{r}
 + \frac{\pi \eps_{10} \alsgt}{60\, C_F \alst} \,r^3 \; ,
\ee
where $\eps_{10} = 1.47$.
Note however that this $H_{\rm eff}$ may not be used for
higher states; an extra nonperturbative piece would have
to be added (see e.g. \cite{bb:leut}).

\section{Charmonium States.}

Only states with $n=1$ can be described in any way with the
rigorous QCD formalism; and even these states can be treated
only approximately. This is certainly to be expected: the average
momentum,
$\bar{k}_{10}$ is of about $0.7$ GeV, so applicability of the
perturbative analysis is a marginal matter.

The mass of the $c$ quark may be found from a formula like that for
the $b$ quark, Eq.~(\ref{eq:massfull}) with appropriate changes. We
will only retain the terms $A_2$, $A_3$, $A_{\rm NP}$. The
criterion
\bdm
\vert A_3 \vert = \vert A_{\rm NP} \vert
\edm
is not applicable here, as it occurs for $\mu \ll 1$ GeV. We will
instead require
\bdm
\vert A_2 \vert = \vert A_{\rm NP} \vert \; .
\edm
We find
\bea
\label{eq:mc}
  m_c &=&  1570
\phantom{a}^{+19}_{-19} \,\,(\La)
\phantom{a}^{-7}_{+7} \,\, (\alsgt) \
{\rm MeV} \; ,
\\
\label{eq:mcbar}
 \mcbar(\mcbar) &=&  1306
\phantom{a}^{+21}_{-34} \,\,(\La)
\phantom{a}^{-6}_{+6} \,\, (\alsgt) \
{\rm MeV} \; ,
\eea
with
\bea
\label{eq:nummc}
 && \mu = 1.00 \phantom{a}^{+0.34}_{-0.24} \;{\rm GeV} \; ,
  \als = 0.36
 \; , \;
 \alst  = 0.53
 \; ,
\\
 && A_2 = -A_{\rm NP}= -0.063
\; ,\;
A_3 = - 0.013
 \; , \;
 \,{\rm for}\; \La(n_f=4,\mbox{2--loop}) = 0.2 \;{\rm GeV}
\nonumber
\eea
(we do not present estimates of systematic errors for the $c\bar{c}$
system as they would be meaningless).
Eq.~(\ref{eq:mcbar}) has to be compared with the standard value
\bdm
 \mcbar(\mcbar) = 1.270 \pm 0.050 \;{\rm GeV}\;.
\edm
The agreement is very good.

For the hyperfine splitting we use the $c$--quark version of
Eq.~(\ref{eq:dmvp}). The criterion that perturbative corrections
balance nonperturbative ones
would give
\be
\label{eq:dmjpetc}
 M(J/\psi)-M(\eta_c) = 70 \;{\rm MeV}
\ee
with, unfortunately
\be
\mu = 0.86 \; ,\;
\de_{\rm hf}^{p} =  0.18 \; , \;
\de_{\rm hf}^{NP}=  4.50 \; , \;
\de_{\rm wf} = -0.57  \; . \;
\ee
The values of $\mu$
and $\de_{\rm hf}^{NP}$
make the QCD analysis unreliable. If we take
$\mu = 1.00$ from (\ref{eq:nummc}), we get $337$ MeV for the
splitting with, unfortunately, $\de_{\rm hf}^{\rm NP} = 2.75
\;(!)$. The best result is obtained with $\mu= 1.00$ GeV and
{\em only} the leading term (i.e., we neglect
$\de_{\rm hf}^{\rm p}$, $\de_{\rm hf}^{\rm NP}$
and $\de_{\rm wf}$).
This gives
\be
 M(J/\psi)-M(\eta_c) = 93 \;{\rm MeV}\;({\rm exp}:\, 118
\; {\rm MeV})\;.
\ee
Finally, for the $e^{+}e^{-}$ decay of $J/\psi$ we have a formula
like
(\ref{eq:gamee}). If we {\em neglect} $\de_r$,
$\de_{\rm hf}^{\rm p}$, $\de_{\rm hf}^{\rm NP}$,
we get the leading result
\bea
\Gamma^{(0)}_{e^{+}e^{-}}(J/\psi) = 2.90 \;{\rm keV}\;,
(\mu = 0.99 \;{\rm GeV}) \;,
\eea
If we impose the criterion that perturbative and nonperturbative
contributions cancel we find instead
\bea
\label{eq:gameejp}
&& \Gamma_{e^{+}e^{-}}(J/\psi) = 2.43 \;{\rm keV}\;,
\;({\rm exp}: \, 5.4 \pm 0.3 \;{\rm keV})\;,
\\
&& \mu = 1.07 \;{\rm GeV} \;,
\de_r = -0.58 \; ,
\de_{\rm wf} = -0.41 \;,
\de_{\rm NP} = 1.64 \;.
\nonumber
\eea
The size of the corrections make this result unreliable.

As is clear from Eqs.~(\ref{eq:dmjpetc}) to (\ref{eq:gameejp}),
only the ground state energy can be
described reliably with QCD. The two other quantities, being
more sensitive to the detail, are only reproduced in order
of magnitude. As could be expected, it is the nonperturbative
effects that are mostly responsible for this breakdown. Indeed,
the wave function extends well into the confinement region
where our treatment of nonperturbative effects (leading order
only) is certainly not a good approximation.

\section{Summary and Conclusions. Connection with
Phenomenological Analyses.}

We present a summary of our results in the form of a table
(Table 1.).
We here also display the experimental figures (when available),
as well as the results of some other existing evaluations.
Because the calculation is not complete, as the
nonperturbative corrections are not known,
we do not present here our (partial) results for LS, tensor,
and hyperfine splittings for the $n=2$ $b\bar b$ system.
The quality of the results may be appreciated by recalling
that we do not have any free parameters (only $\mup$ is
taken from experiment) and no input other than rigorous
QCD has been used. This should be compared with the six
to eight (depending how one counts) parameters of
standard, phenomenological evaluations
(Refs~\cite{bb:gupt,bb:pant}) for example) not to mention the
arbitrariness in the device of the "confining"
potential\footnote{Of course these phenomenological evaluations
predict parts of the spectrum ($n \geq 2$) which are not
accessible to our analysis.} which indeed
one can show to be incorrect for small $r$, cf.
Eq.~(\ref{eq:hameff}).

Besides the energy levels and decay widths we have given in
Eqs.~(\ref{eq:wfup}), (\ref{eq:hameff})
a wave function and an effective Hamiltonian
applicable to the $n=1,\,l=0$ state of heavy $q \bar q$ systems.
The last will serve us as a starting point in the discussion of the
connection between our analysis and standard phenomenological ones.
Eq.~(\ref{eq:wfup}) contains the nonperturbative piece
\be
\label{eq:phennp}
 \frac{\pi \eps_{10} \alsgt}{60 \, C_F \alst}\,r^3 \;,
\ee
whose behaviour for $r \ra \infty$ is very different from the
expected linear increase,
\be
\label{eq:linpot}
 K\,r \; , \; K^{1/2} \sim 0.4 \;{\rm GeV} \;.
\ee
It is obvious
that (\ref{eq:phennp}) has to be considered as only valid in the
region of small $r$, $r \infap a = 2/(m C_F \alst)$.
It is also obvious that the radiative correction,
\bdm
 - \frac{C_F \bt_0 \als(\mu^2)}{2\pi}\,\frac{\log r\mu}{r} \;,
\edm
with $\mu \sim 1/a$ will break down for $r \gg a$.
This means that, in order to be able to describe high
excited states ($n > 1 \;{\rm for}\; c\bar{c}$,
$n > 2 \;{\rm for}\; b\bar{b}$) we must first, modify
(\ref{eq:hameff}) and second, introduce effective confinement
potentials that reproduce the expected large $r$ features, as
is done in
refs.~\cite{bb:gupt,bb:pant}.
In this respect we should mention that the
{\em relativistic} corrections are small, of order
\bdm
 < \!\!{\vec{v}\,}^2\!\!>_{nl} \;=\;
\frac{ (m C_F \alst)^2}{4 \,n^2}\; ,
\edm
so a description in terms of potentials should be meaningful.
Perhaps our results would be useful in joining more realistically
the regions $r \ll a$, $r\gg a$, and even eliminating some of the
parameters of potential models. Applications to $t \bar{t}$ should
be important and are under way.

We finish this paper with a few words on
possible improvements. First of all, one should finish the
calculation of the leading nonperturbative corrections to
hyperfine, LS, and tensor splittings for $n=2$, which would
nicely complete the rigorous QCD evaluation of the $n=1\, ,2$
$b\bar{b}$ system. Secondly, a two--loop (not difficult but
lengthy) evaluation of the spin--independent potential would
allow a reduction of the {\em systematic} error for $m_b$,
$\mbbar(\mbbar^2)$ so that, e.g. the second would become
known to within a dozen or so MeV; work along these lines
is in progress. Finally, better knowledge of the basic parameters
$\La$, $\alsgt$ would certainly be very welcome, and
would substantially
improve our results.

\newpage
\renewcommand{\arraystretch}{1.5}
\begin{table}[b]
\def\MeV{\mbox{\footnotesize\rm{MeV}}}
\def\keV{\mbox{\footnotesize\rm{keV}}}
\begin{center}
\begin{tabular}{|@{}c@{}|c|@{}c@{}|c|@{}c@{}|@{}c@{}|}
\hline
Quantity & Refs.~2 & Ref.~4 & Refs.~5,9,11 & This Work & Experiment
\\
\hline
  $m_b$(pole)
& $4780$
& $\,4840$ to $4880\,$
&
& $\,\, 4906\phantom{a}^{+69}_{-51}\phantom{a}^{-4}_{+4}
  \phantom{a}^{+11}_{-40} \;{\MeV}\,$
&
\\
\hline
  $\bar{m}_b(\bar{m}_b^2)$
&
&
& $4250 \pm 100$
& $ 4397\phantom{a}^{+7}_{-2}\phantom{a}^{-3}_{+4}
  \phantom{a}^{+16}_{-32} \;{\MeV}$
&
\\
\hline
  $\scriptstyle{M(\Upsilon) - M(\eta_b)}$
& $35$
& $30$ to $46$
& $110$
& $ 36\phantom{a}^{+13}_{-7}\phantom{a}^{+3}_{-6}
 \phantom{a}^{+11}_{-5}
  \;{\MeV}$
&
\\
\hline
  $\Gamma_{e^{+}e^{-}}(\Upsilon)$
& $1.29$
& $0.94$ to $1.13$
&
& $ 1.01\phantom{a}^{+0.02}_{-0.02}\phantom{a}^{+0.18}_{-0.24}
  \;{\keV}$
& $\, 1.34 \pm 0.04$ \keV\,
\\
\hline
  $\Gamma_{2\gamma}(\eta_b)$
&
&
&
& $0.17$ \keV
&
\\
\hline
  $\,\scriptstyle{M(\Upsilon)_{2S}-M(\Upsilon)_{1S\,}}$
& $551$
&
&
& $ 479\phantom{a}^{+28}_{-23}\phantom{a}^{+49}_{-69}
  \;{\MeV}$
& $563$ \MeV
\\
\hline
  $\,\scriptstyle{M(\Upsilon)_{2P}-M(\Upsilon)_{1S\,}}$
& $431$
&
&
& $ 417\phantom{a}^{+25}_{-20}\phantom{a}^{+42}_{-59}
  \;{\MeV}$
& $428$ \MeV
\\
\hline
  $\,\scriptstyle{M(\Upsilon)_{2P}-M(\Upsilon)_{2S\,}}$
& $120$
&
&
& $128$ to $214$
& $135$ \MeV
\\
\hline
  $m_c$(pole)
& $1200$
& $1480$
&
& $\,\, 1570\phantom{a}^{+19}_{-19}\phantom{a}^{-7}_{+7}
   \;{\MeV}$
&
\\
\hline
  $\bar{m}_c(\bar{m}_c^2)$
&
&
& $1270 \pm 50$
& $\,\, 1306\phantom{a}^{+21}_{-34}\phantom{a}^{-6}_{+6}
   \;{\MeV}$
&
\\
\hline
  $\scriptstyle{M(J/\psi) - M(\eta_c)}$
& $116$
&
&
&
& $118$ \MeV
\\
\hline
  $\Gamma_{e^{+}e^{-}}(J/\psi)$
& $5.07$
& $4.5$
&
&
& $5.4\pm 0.3$ \keV
\\
\hline
  $\Gamma_{2\gamma}(\eta_c)$
&
&
&
& $3$
& $6 \phantom{a}^{+6}_{-5}$ \keV
\\
\hline
\end{tabular}
\end{center}
\caption{Compilation of results.}
\vskip0.6cm
\vbox{
\indent
In what refers to our results in this work, the first error is due
to varying
$\La(\mbox{2--loop},n_f=4) = 200 \phantom{a}^{+80}_{-60}
\;{\rm MeV}$, the second to
$ \alsgt = 0.042 \pm 0.020 \; {\rm GeV}^4$;
the third error, when given, is an estimated systematic error.
}
\end{table}
\renewcommand{\arraystretch}{1.0}

\newpage
\appendix{
\large{{\bf Appendix I.\, Fourier Transforms.}}
}
\vskip0.4cm
\begin{eqnarray*}
&\displaystyle{
 \Vt(\vk)
\phantom{aaaaaaaaaaaaaaaa}} &
V(\vr) = (2\pi)^{-3} \,\int\! d^3\! k \,e^{i \vk \cdot \vr}\,\Vt(\vk)
\\
&&
\\
&\displaystyle{
 \log k
\phantom{aaaaaaaaaaaaaaaa}} &
 - \frac{1}{4\pi} {\rm reg}\,\frac{1}{r^3}
\\
&\displaystyle{
 \frac{1}{k}
\phantom{aaaaaaaaaaaaaaaa}} &
\frac{1}{2\pi^2}\,\frac{1}{r^2}
\\
&\displaystyle{
\frac{\log k}{k}
\phantom{aaaaaaaaaaaaaaaa}} &
- \frac{1}{2\pi^2}\,\frac{\log r + \eul}{r^2}
\\
&\displaystyle{
\frac{1}{k^2}
\phantom{aaaaaaaaaaaaaaaa}} &
\frac{1}{4 \pi}\,\frac{1}{r}
\\
&\displaystyle{
\frac{\log k}{k^2}
\phantom{aaaaaaaaaaaaaaaa}} &
- \frac{1}{4\pi}\,\frac{\log r + \eul}{r}
\\
&\displaystyle{
\frac{\log^2 k}{k^2}
\phantom{aaaaaaaaaaaaaaaa}} &
\frac{1}{4\pi}\,\frac{(\log r + \eul)^2
+ \pi^2/12}{r}
\\
&\displaystyle{
\Lak
\phantom{aaaaaaaaaaaaaaaa}} &
\frac{1}{4\pi r^3} \vS \cdot \vL
\\
&\displaystyle{
\Lak\log k
\phantom{aaaaaaaaaaaaaaaa}} &
\frac{1- \log r}{4\pi r^3} \vS \cdot \vL
\\
&\displaystyle{
 \frac{1}{3} T(\vk)
\phantom{aaaaaaaaaaaaaaaa}} &
\frac{1}{4\pi r^3}\,S_{12}(\vr)
\\
&\displaystyle{
 \frac{\log k}{3} \,T(\vk)
\phantom{aaaaaaaaaaaaaaaa}} &
\frac{4/3 - \log r}{4\pi r^3}
\,S_{12}(\vr)
\end{eqnarray*}
Here
\begin{eqnarray*}
\int d^n r \phi(\vr) \,{\rm reg}\,\frac{1}{r^n}
 &\equiv&  \lim_{\eps \to 0}
\left\{
 \int\!d^n r \,\phi(\vr)\,\frac{r^{\eps}}{r^n} - A(n,\eps)\phi(0)
\right\}
\;,
\\
A(n,\eps)
 &\equiv&
 \frac{2\pi^{n/2}}{\Gamma(n/2)}
\left\{
 \frac{1}{\eps} + \log 2 + \frac{\psi(n/2) - \eul}{2}
\right\}
\;.
\end{eqnarray*}
Moreover,
\begin{eqnarray*}
\La
 &\equiv&
- i \vS \cdot \frac{\vk \times \vp}{k^2} \; , \;
\; T  \equiv
\frac{\vk^2 \vsi_1 \cdot \vsi_2 -
3(\vk \cdot \vsi_1) (\vk \cdot \vsi_2)}{k^2} \; ,
\\
 S_{12} &\equiv&  \frac{3}{r^2}(
(\vr \cdot \vsi_1) (\vr \cdot \vsi_2) - (\vsi_1 \cdot \vsi_2) \, = \,
2 \sum_{i j}(\frac{3}{r^2} r_i r_j -\de_{ij}) S_i S_j \; ;
\\
 \vS & = & \frac{1}{2}(\vsi_1 + \vsi_2) \; , \;
\;\vL = \vr \times \vp \;.
\end{eqnarray*}

\newpage
\appendix{
\large{{\bf Appendix II. \, Expectation Values.}}
}
\vskip0.4cm
We present here a few expectation values in (unperturbed)
Coulombic wave functions for hydrogen--like systems, as some of
them cannot be found in the standard literature. We define $a$ as
the Bohr--like radius; i.e., for a potential $-\beta/r$, and a
particle with reduced mass $m_r$, $a= 1/m_r \beta$. We have:
\begin{eqnarray*}
 <\!{\rm reg}\,\frac{1}{r^3}\!>_{n0}
 & = &
 \frac{4}{n^3 a^3}\left\{ \log \frac{n a}{2} -
 \sum_{k=1}^{n} \frac{1}{k} -\frac{n-1}{2 n} \right\} \;,
\\
 <\!{\rm reg}\,\frac{1}{r^3}\!>_{nl}
 & = &
 \frac{2}{n^3 a^3} \,\frac{1}{l(l+1)(2 l+1)} \;, \; l \neq 0 \;
\\
 <\!\frac{\log r}{r}\!>_{nl}
 & = &
\frac{\log n a/2 + \psi(n+l+1)}{n^2 a}
\\
 <\!\frac{\log r}{r^2}\!>_{nl}
 & = &
\frac{2}{n^3 (2 l+1) a^2}
 \left\{ \log \frac{n a}{2} - \psi(n+l+1)
+ \psi(2 l+2) + \psi(2 l+1) \right\}
\\
 <\!\frac{\log r}{r^3}\!>_{nl}
 & = &
\frac{2}{n^3 l(l+1)(2 l +1) a^3}
\\
 && \ \ \ \ \ \times
 \left\{ \log \frac{n a}{2} - \psi(n+l+1)
+ \psi(2 l+3) + \psi(2 l)-\frac{n-l-1/2}{n} \right\} \; , \;
 l \neq 0
\\
 <\!\frac{\log^2 r}{r}\!>_{10}
 & = &
 \frac{1}{a} \left\{
 \log^2 \frac{a}{2} + 2(1 - \eul) \log \frac{a}{2} + (1-\eul)^2 +
 \frac{\pi^2}{6} - 1 \right\}
\\
 <\!\frac{1}{r}\Delta\!>_{nl}
 & = &
 \frac{2 l +1 - 4 n}{n^4(2 l +1)a^3}
\\
 <\!\frac{\log r}{r}\Delta\!>_{nl}
 & = &
 \frac{1}{n^4(2 l +1)a^3}
 \left\{ (2 l +1 - 4 n)\log\frac{n a}{2}
 \right.
\\
&& \ \ \ \ \ \ \
 \left.
+ (2 l +1 + 4 n)\psi(n+l+1)
 - 4 n \left[ \psi(2l+2) +\psi(2l+1)\right]
 \phantom{\frac{1}{2}} \!\!\!\!\! \right\} \;.
\end{eqnarray*}
\vskip0.6cm
\noindent
Here
\bdm
 \psi(x) = \frac{{\rm dlog}\,\Gamma(x)}{{\rm d}x}
\edm
The formulas involving logarithms may be obtained
by differentiating the following expressions:
\begin{eqnarray*}
 && <\! r^p\! > \,=\, a^p \,\frac{n^{p-1}}{2^{1+p}}\,
\frac{(n-l-1)!}{(n+l)!}
\\
&& \ \ \ \ \ \ \times \,\left\{
\begin{array}{ll}
 \displaystyle{
 \sum_{r=0}^{n-l-1} \, \frac{\Gamma(2l+3+p+r)\Gamma^2(2+p)}
{\Gamma(r+1)\Gamma^2(n-l-r)\Gamma^2(3+l+p-n+r) }}
 &\ \ , \, p > -2\;,
\\
 \displaystyle{
 \sum_{r=0}^{n-l-1} \, \frac{\Gamma(2l+3+p+r)\Gamma^2(n-l-2-p-r)}
{\Gamma(r+1)\Gamma^2(n-l-r)\Gamma(-1-p) }}
 &\ \ , \, p < -1\;,
\end{array}
\right.
\end{eqnarray*}
\newpage
These expressions may be derived from the corresponding
ones for Laguerre polynomials
\begin{eqnarray*}
 &&
\int^{\infty}_{0} \!dx\, x^\la e^{-x}\,
\left[ L^{k}_{\nu}(x)\right]^2
\\
&& \ \ \ \ \ \ = \,\left\{
\begin{array}{ll}
 \displaystyle{
 \sum_{r=0}^{\nu} \, \frac{\Gamma(\la+1+r)\Gamma^2(\la+1-k)}
{\Gamma(r+1)\Gamma^2(\nu+1-r)\Gamma^2(\la+1-k-\nu+r)}}
 &\ \ , \, \la > k-1\;,
\\
 \displaystyle{
 \sum_{r=0}^{\nu} \, \frac{\Gamma(\la+1+r)\Gamma^2(\nu+k-\la-r)}
{\Gamma(r+1)\Gamma^2(\nu+1-r)\Gamma^2(k-\la) }}
 &\ \ ,\, \la < k\;,
\end{array}
\right.
\end{eqnarray*}
and these in turn are obtainable from the Schr\"{o}dinger
formulas of Ref.~\cite{bb:beth} (eq.~(3.11) page 13) and
Ref.~\cite{bb:gali} (Appendix~A, eq.~A-80).

\newpage
{\large {\bf List of figures.}}
\vskip0.4cm
\leftline{Figure 1. \phantom{aa}
One--loop diagrams for the $q \bar q$
system.}
\vskip0.2cm
\leftline{Figure 2. \phantom{aa}
Kinematics for the tree diagram in the $q\bar{q}$ system.}
\vskip0.2cm
\leftline{Figure 3. \phantom{aa}
Gluon condensate contributions to $q\bar{q}$. }
%
%
%
%
%

%
%
\end{document}